\documentclass[aps,nofootinbib,preprint]{revtex4}
\usepackage{graphicx}
\usepackage{slashed}

\def\ga{\mathrel{\raise.3ex\hbox{$>$\kern-.75em\lower1ex\hbox{$\sim$}}}}
\def\la{\mathrel{\raise.3ex\hbox{$<$\kern-.75em\lower1ex\hbox{$\sim$}}}}

\begin{document}

\title{Two Higgs Doublet Model in light of the Standard Model 
$H \to \tau^+ \tau^-$ search at the LHC}
\author{Abdesslam Arhrib~\footnote{aarhrib@ictp.it}}
\affiliation{D\'epartement de Math\'ematiques, Facult\'e des Sciences et
  Techniques, Tanger, Morocco}
\affiliation{Department of Physics, National Cheng Kung
  University, Taiwan 701, Taiwan}
\affiliation{Laboratoire de Physique des Hautes Energies 
et Astrophysique, D\'epartement de Physiques, Facult\'e des Sciences Semlalia, Marrakech, Morocco}

\author{Cheng-Wei Chiang~\footnote{chengwei@ncu.edu.tw}}
\affiliation{Department of Physics and Center for Mathematics and Theoretical Physics,
National Central University, Chungli 32001, Taiwan}
\affiliation{Institute of Physics, Academia Sinica, Taipei 11529, Taiwan}
\affiliation{Physics Division, National Center for Theoretical Sciences, Hsinchu 30013, Taiwan}

\author{Dilip Kumar Ghosh~\footnote{tpdkg@iacs.res.in}}
\affiliation{Department of Theoretical Physics, Indian Association for the Cultivation of Science, 2A \& 2B Raja S.C. Mullick Road, Kolkata 700 032, India}

\author{Rui Santos~\footnote{rsantos@cii.fc.ul.pt}}
\affiliation{Instituto Superior de Engenharia de Lisboa, Rua Conselheiro Em\'\i dio Navarro 1, 1959-007 Lisboa, Portugal,}
\affiliation{Centro de F\'\i sica Te\' orica e Computacional, Faculdade de Ci\^encias, Universidade de Lisboa
Av. Prof. Gama Pinto 2, 1649-003 Lisboa, Portugal.}

\begin{abstract}
We study the implications of the searches based on
$H \to \tau^+ \tau^-$ by the ATLAS and CMS collaborations
on the parameter space of the two Higgs doublet model (2HDM).
In the 2HDM, the scalars can decay into a tau pair with a branching
ratio larger than the SM one, leading to constraints on the 2HDM parameter space. 
We show  that in model II, values of $\tan \beta >1.8 $ are definitively
excluded if the pseudo-scalar is in the mass range $110 \, GeV < m_A < 145 \, GeV$.
We have also discussed the implications for the 2HDM
of the recent di-muon search by the ATLAS collaboration
for a CP-odd scalar in the mass range 4-12 GeV.
\end{abstract}

\maketitle

\newpage

\section{Introduction}
The Standard Model (SM) of particle physics has been very 
successful in explaining most of 
the experimental observations in elementary particle phenomena. 
The most crucial and only missing ingredient of the 
electroweak sector of the SM, the Higgs boson
\cite{Higgs:1964ia,Englert:1964et}, has finally
been hinted at CERN's Large Hadron Collider (LHC) by the 
ATLAS~\cite{ATLAS-CONF-2011-161, ATLAS-CONF-2011-163} 
and CMS~\cite{CMS-PAS-HIG-11-030, CMS-PAS-HIG-11-032} collaborations. A slight
excess was measured both in the $\gamma \gamma$ and in the $WW^{*}, ZZ^{*}$ 
channels, which is compatible with a SM Higgs boson with a mass of about 125 GeV.


Even if a discovery cannot still be claimed, there are well 
established limits for the SM Higgs. Unitarity of longitudinal weak gauge
boson scattering at high energies leads to the bound $M_H \alt 700$ 
GeV \cite{Lee:1977eg}. Direct searches of the SM Higgs boson at 
LEP2 has set a lower bound on the mass $m_H \agt 114.4$
GeV at $95\%$ confidence level (CL) \cite{Barate:2003sz}.  
The CDF and D0 Collaborations
at the Tevatron have excluded the Higgs boson mass in the 
range $156-177$ GeV at $95\%$ CL 
\cite{Aaltonen:2010cm,Abazov:2010ct,Aaltonen:2011gs}.  
The ATLAS and CMS Collaborations have searched for the SM 
Higgs boson in the $H \to WW $ channel.  
The ATLAS Collaboration has excluded the mass range 
$\sim 155-190$ GeV and $\sim 295- 450$ GeV at $95\%$ CL 
\cite{ATLAS_Higgs}, while the CMS Collaboration has 
ruled out the Higgs boson mass
 in the range of $\sim 149-206$ GeV and $\sim 300-400$ GeV at 
$95\% $ CL \cite{CMS_Higgs}. 
 Very recently, both ATLAS and CMS Collaborations have 
updated their results on the Higgs boson 
 searches and excluded $m_H$ in the mass ranges 
[112.7 ,115.5] GeV, [131,237] GeV and 
[251,453] GeV for ATLAS \cite{ATLAS-CONF-2011-163} and 
[128,525] GeV for CMS \cite{CMS-PAS-HIG-11-032}, at $95\%$ CL.  
 

In the SM, a light Higgs decays into a $\tau^+\tau^-$ pair with a 
branching ratio $\sim 10\%$, which makes it a promising search channel.  
This channel suffers less from QCD background than the more dominant 
$b\bar b$ channel at the LHC.  Moreover, in some versions of 
models such as the two Higgs doublet model (2HDM),
the couplings of the Higgs boson to the third-generation fermions can be
strongly enhanced over a large portion of the parameter space; 
{\it i.e.}, when $\tan\beta$ is large ($\tan \beta = v_2 /v_1$ is the ratio of the two 
vacuum expectation values (VEV's)). It is therefore useful and 
timely to constrain such models using this channel. A study
for the MSSM using this channel was recently performed in~\cite{Baglio:2011xz}. 
Searches for Higgs bosons decaying to tau pairs that in turn decay into final
states with one or two light leptons at the LHC have been performed by CMS
\cite{CMS-PAS-HIG-11-029} and ATLAS \cite{Aad:2011rv} at the CM 
energy of $7$ TeV.  Even though the sensitivity of ATLAS and CMS 
searches has not yet reached the SM sensitivity, useful constraints 
can be put on new physics model.  
It is the purpose of this paper to study the implications of 
CMS and ATLAS data on the parameters of the 2HDM. 

In fact, there are several scenarios in the 2HDM, differing mainly in 
how the Higgs bosons couple to the SM fermions. One common trend in 
all scenarios is that the Higgs boson spectrum is the same after 
electroweak symmetry breaking (EWSB): two charged Higgs particles 
$H^\pm$, two CP-even scalars, $H, h$ and one CP-odd scalar, $A$. By limiting 
ourselves to natural flavor conservation (NFC), the 
Higgs couplings to the fermions are completely fixed by $\alpha$ and 
$\tan\beta$, where $\alpha$ is the mixing angle in the CP-even sector 
and $\tan \beta$.  
Therefore, the production rate and decay branching 
ratios of the Higgs bosons in this model can be worked out readily.

In this paper, we investigate the predictions for
the $pp (gg + b\bar{b}) \to h (A) \to \tau^+\tau^-$ cross sections and examine
the consequences for the 2HDM parameter space. We also discuss the very light CP-odd
scenario $m_A\sim 4-12$ GeV which decays mainly to a pair of muons for which
there are some preliminary results from the ATLAS Collaboration \cite{ATLASmumu}.

This paper is organized as follow.  In Section~\ref{sec:yukawa} 
we introduce the patterns of the Yukawa interactions of the 2HDM, while in  
Section~\ref{sec:results} we illustrate our results for CP-even 
and CP-odd Higgs bosons which decay to a pair of tau leptons.  A 
discussion of a very light CP-odd scalar decaying to a pair of muons 
is presented in Section~\ref{sec:light}.
 We summarize our findings in Section~\ref{sec:summary}.

\section{2HDM and the patterns of the Yukawa interactions \label{sec:yukawa}}
There are several alternatives to the SM that have started being tested at the
LHC. The most popular ones are the Minimal Supersymmetric Standard 
Model (MSSM) or some of its scalar sector variants like the 2HDMs, 
little Higgs models and extra dimensions models among others. 
The 2HDMs are formed by adding an extra complex 
$SU_L(2)\times U(1)_Y$ scalar doublet with a non-vanishing vacuum expectation
value (VEV) to the SM Lagrangian. In the CP-conserving version of 2HDM, 
we end up with two CP-even  scalars, $h$ and $H$, one CP-odd scalar $A$ 
and a pair of charged Higgs bosons, $H^\pm$.

The Yukawa Lagrangian of the 2HDM is a straightforward 
generalization of the SM one. However, flavor changing 
neutral currents (FCNC) may occur at tree-level. To avoid tree-level FCNC,
it suffices that fermions of a given electric charge couple to no more 
than one Higgs doublet~\cite{Glashow:1976nt}. 
It is known that there are four patterns of the Yukawa interaction, free
from tree-level FCNC, 
depending on the assignment of charges for quarks and
leptons under the $Z_2$ symmetry 
($\phi_1 \to \phi_1; \, \phi_2 \to - \phi_2$) 
\cite{Barger:1989fj,Grossman:1994jb,Akeroyd:1994ga}.
Hereafter, we define as Type I the model where only the doublet $\phi_2$
couples to all fermions; Type II is the model where $\phi_2$ couples to
up-type quarks and $\phi_1$ couples to down-type quarks and charged 
leptons; in Type
III model $\phi_2$ couples to up-type quarks
and charged leptons and $\phi_1$ couples to down-type quarks;
the Type IV model is instead built such that $\phi_2$ couples to all 
quarks and $\phi_1$ couples to all leptons. Models III and IV
have been explored in a number of 
papers~\cite{Barnett:1984zy,Barnett:1983mm,Goh:2009wg} and more recently  with
a renewed interest in~\cite{Su:2009fz,Logan:2009uf,Aoki:2009ha}. 

\begin{table}[h]
\begin{center}
\begin{tabular}{|c|c|c|c|c|c|c|}
\hline
& $y_h^u$ & $y_h^d$ & $y_h^\ell$
& $y_A^u$ & $y_A^d$ & $y_A^\ell$ \\ \hline
Type-I
& {$\frac{c_\alpha}{s_\beta}$} & $\frac{c_\alpha}{s_\beta}$ & $\frac{c_\alpha}{s_\beta}$
& $\cot\beta$ & $-\cot\beta$ & $-\cot\beta$ \\
Type-II
& $\frac{c_\alpha}{s_\beta}$ & $-\frac{s_\alpha}{c_\beta}$ & 
$-\frac{s_\alpha}{c_\beta}$
& $\cot\beta$ & $\tan\beta$ & $\tan\beta$ \\
Type-III
& $\frac{c_\alpha}{s_\beta}$ & $-\frac{s_\alpha}{c_\beta}$ & $\frac{c_\alpha}{s_\beta}$
& $\cot\beta$ & $\tan\beta$ & $-\cot\beta$ \\
Type-IV
& $\frac{c_\alpha}{s_\beta}$ & $\frac{c_\alpha}{s_\beta}$ & $-\frac{s_\alpha}{c_\beta}$
& $\cot\beta$ & $-\cot\beta$ & $\tan\beta$ \\
\hline
\end{tabular}
\end{center}
\caption{Yukawa couplings to the scalars $h$ and $A$ normalized to the 
 SM Higgs Yukawa couplings according to Eq.~(\ref{Eq:Yukawa}).} 
\label{Tab:MixFactor}
\end{table}

After EWSB, the Yukawa interactions of the neutral scalars $h$ and $A$ 
are expressed in terms of 
mass eigenstates  as:
\begin{eqnarray}
\label{Eq:Yukawa}
{\mathcal L}_{Y}^{2HDM} =
&& - \frac{gm_f}{2 m_W} y_h^f{\overline f}f h 
-i\frac{gm_f}{2 m_W} y_A^f{\overline f}\gamma_5f A
\end{eqnarray}
where the factors $y^f_{h,A}$, needed for this study, 
are given in Table~\ref{Tab:MixFactor} in the four Yukawa types of 
the 2HDM models.

As can be seen from Table~\ref{Tab:MixFactor}, the models can be grouped 
in two pairs: (I, IV) and (II, III), as the only difference within each 
of these two pairs is in the Higgs couplings to the charged leptons. 
Moreover, the way we have chosen to build the Yukawa Lagrangian is such that
up-type quarks have the same couplings to the Higgs bosons in all four
types. Down-type quarks have different couplings in the two groups defined
above, (I, IV) and (II, III), as can be seen in 
Table~\ref{Tab:MixFactor}. 
In the large $\tan\beta$ limit, in 2HDM-(II, III), 
the Higgs coupling to a pair of down-type
quarks is enhanced by a factor of $1/\cos\beta \approx \tan\beta \, 
(\tan \beta \gg 1) $, while for 2HDM-(I, IV) there is no such enhancement. 

It is well-known that 2HDM parameters can be constrained both 
from theory and experiment. For example, 
from perturbativity arguments, the top and bottom Yukawa couplings,
$Y_{t,b}=(\frac{g}{\sqrt{2}m_W} \{ m_t \cot\beta,m_b \tan\beta\}$,
 cannot be too large. Therefore, 
the requirement that $|Y_{t,b}|^2<4\pi$ at the tree-level provides a 
constraint on $\tan\beta$ which reads $0.1\lesssim\tan\beta\lesssim 100$. 
Tree-level unitarity~\cite{Kanemura:1993hm, Akeroyd:2000wc, Horejsi:2005da}
and vacuum stability~\cite{Deshpande:1977rw} also impose severe
constraints on the 2HDM parameter space~\cite{Aoki:2011wd}. We note
that once a CP-conserving minimum is chosen, the 2HDM is naturally
protected against charge and CP-breaking~\cite{Ferreira:2004yd, Maniatis:2006fs, Ivanov:2006yq}.

Experimental data and in particular precision data, are now
accurate enough to put constraints 
on new physics models such as the 2HDM's. In this regard, 
the 2HDM is subject to a number of constraints
from which $b\to s\gamma$, 
$Z\to b\bar{b}$, $\delta\rho$, $g-2$ and $B\to l \nu$~\cite{WahabElKaffas:2007xd,Mahmoudi:2009zx,Cheung:2003pw,
Cheung:2001hz, Haber:1999zh} are the most relevant. Some of 
those constraints, such as the ones related to B physics
observables, put already some severe constraints on the charged 
Higgs boson mass and on 
$\tan\beta$~\cite{WahabElKaffas:2007xd,Mahmoudi:2009zx}. 
Other constraints like $Z\to b\bar{b}$~\cite{Haber:1999zh} 
and the muon $g-2$~\cite{Cheung:2003pw,Cheung:2001hz} can restrict the available
parameter space in the neutral Higgs sector as well.
All the above constraints
will be taken into account in our analysis.

\section{Numerical Results \label{sec:results}}
Before presenting our results, we will discuss the direct
experimental constraints on the mass of the lightest CP-even Higgs boson
and on the mass of the CP-odd scalar.
For the neutral Higgs bosons, the experimental searches depend on 
the mixing parameters $\alpha$ and $\tan\beta$ and on their decay patterns. 
If the neutral Higgs bosons decay mainly into fermions, 
the OPAL, DELPHI and L3 collaborations have set a limit on the masses of
$h$ and $A$ in the 2HDM~\cite{Abbiendi:2004gn,Abdallah:2004wy}.
OPAL claims an exclusion almost independent of the values of 
$\alpha$ and $\tan\beta$,  in the mass ranges
$1\la m_{h} \la 55$ GeV and $3\la m_{A} \la 63$ GeV~\cite{Abbiendi:2004gn}.
It should be noted however that there are several scenarios
where the above limits do not hold. First, if either $m_h$ or $m_A$ 
are above the kinematical threshold in either of the production
processes $e^+ e^- \to hZ$ or $e^+ e^- \to hA$, no bound
on the masses can be derived. Second, if $\sin (\alpha - \beta) \approx 0$, 
$\sigma_{e^+ e^- \to hZ} \approx 0$ and again the
bound on the CP-even mass do not hold.  On the other hand,
if $\sin (\alpha - \beta) \approx 1$, 
$\sigma_{e^+ e^- \to h A} \approx 0$ and in that
scenario no bound on the CP-odd mass can be extracted.
Assuming that the Higgs-strahlung cross section is the SM one , 
L3 sets a lower limit on $m_h$ of 
about 110.3 GeV \cite{Abbiendi:2004gn}. Therefore, we will assume from now on
that both $m_h$ and $m_A$ are 
in the range 110--140 GeV except for a short section where we 
discuss the case of a very light CP-odd scalar.
Regarding the charged Higgs boson, the combined null searches from all 
four CERN LEP collaborations imply the lower limit 
$M_{H^+}> 78.6$ GeV (95\% CL), a limit which applies to all models in which 
$Br(H^\pm \to \tau \nu )+ Br(H^\pm \to c\bar{s})=1$ \cite{Schael:2006cr}.

In order to extract limits on the 2HDM parameters,
we will focus on the following observables
%
%
\begin{eqnarray}
& R_{\sigma} &= \frac{\sigma(gg+b\bar{b}\rightarrow \Phi)^{2HDM}}
{\sigma(gg+b\bar{b}\rightarrow \Phi)^{SM}} \, , \\   
& R_{br} & = \frac{Br(\Phi \rightarrow \tau\tau)^{2HDM}}
{Br(\Phi\rightarrow \tau\tau)^{SM}}\, , \\
& R_{\tau\tau} & =  R_{\sigma} \times R_{br} ~,
\end{eqnarray}
where $\sigma(gg+b\bar{b}\rightarrow \Phi)$ is the production 
cross section of the scalar particle
which has been evaluated at next-to-leading-order (NLO) using
HIGLU~\cite{Spira:1995mt} and bb@nnlo~\cite{Harlander:2003ai} with
the CTEQ6~\cite{Nadolsky:2008zw} parton distribution function.
Note that we include the $b\bar{b}$ fusion contribution since in the 
2HDM the $b\bar{b}\phi$ coupling can substantially enhance the 
production cross section. 
We will then use the limits obtained by the CMS~\cite{CMS-PAS-HIG-11-029} 
Collaboration for $R_{\tau\tau}$ at 95\% CL.

The decay widths of the scalar particles 
$\Phi = h, A$ are computed at leading order, both in the SM and in the 2HDM's as
\begin{eqnarray}
\Gamma_\Phi=\sum_{f=\tau,b,c}\Gamma(\Phi\to ff) + \Gamma(\Phi\to WW^*)+
\Gamma(\Phi\to ZZ^*) + \Gamma(\Phi\to gg)+\Gamma(\Phi\to \gamma\gamma)
\end{eqnarray}                            
where the expressions for the scalars decay widths and the respective
QCD corrections are taken from 
\cite{Djouadi:2005gi}.
We note that when $\Phi=A$, the decays  
$\Phi\to \{WW^*,ZZ^*\}$ are not allowed.

\subsection{CP-odd case}

The branching ratio of SM Higgs into $\tau^+ \tau^-$ is of the order of 8\%
for a 100 GeV Higgs boson and drops quickly with an increasing mass.
In the 2HDM, $\Phi\to \tau^+ \tau^-$ mat be enhanced at large 
 $\tan\beta$, for some types of the 2HDMs under consideration.
Such an enhancement can be very large relative to the SM, reaching $\sim30 \%$
in Type II or even $100 \%$ in Type IV.

We start by discussing the case 
of the CP-odd scalar in the 2HDM.
Due to CP invariance, $A$ does not
couple to $WW$ and $ZZ$ and the partial decay widths
to  $gg$ and $\gamma \gamma$ are very small. Considering
that the channel $A \to h Z$ is closed, the CP-odd scalar
can only decay into fermions. We have then
\begin{eqnarray}
Br(A \to \tau \tau)=\frac{\Gamma(\Phi\to \tau\tau)}{\sum_{f=\tau,b,c}
\Gamma(\Phi\to ff)} ~.
\end{eqnarray}
Depending on the Yukawa structure, the above ratio will take different forms
which are independent of $m_A$ in the mass region under consideration and
can be written as
\begin{eqnarray}
Br(A\to \tau \tau)^{I}&=&\frac{m_\tau^2 \cot\beta^2}{m_\tau^2 \cot\beta^2+ 3
  m_b^2 \cot\beta^2+ 3 m_c^2 \cot\beta^2}\approx 0.075 \\
Br(A \to \tau \tau)^{II}&=&\frac{m_\tau^2 \tan\beta^2}{m_\tau^2 \tan\beta^2+ 3
    m_b^2 \tan\beta^2+ 3 m_c^2 \cot\beta^2}\approx 
\left\lbrace
\begin{array}{l}
0.1 \ \mbox{ for }  \tan\beta = 100\\
0.075 \ \mbox{ for } \tan\beta =1
\end{array}
\right. \nonumber \\
Br(A\to \tau \tau)^{III}&=&\frac{m_\tau^2 \cot\beta^2}{m_\tau^2 \cot\beta^2+ 3
    m_b^2 \tan\beta^2+ 3 m_c^2 \cot\beta^2}\approx 
\left\lbrace
\begin{array}{l}
10^{-8} \ \mbox{ for} \tan\beta = 100 \\
0.075  \ \mbox{ for } \tan\beta = 1
\end{array}
\right.\nonumber\\
Br(A \to \tau \tau)^{IV}&=&\frac{m_\tau^2 \tan\beta^2}{m_\tau^2 \tan\beta^2+ 3
    m_b^2 \cot\beta^2+ 3 m_c^2 \cot\beta^2}\approx 
\left\lbrace
\begin{array}{l}
1 \ \mbox{ for} \tan\beta =100\\
0.075 \ \mbox{ for} \tan\beta =1
\end{array}
\right.\nonumber
\end{eqnarray}

\begin{figure}[h!]
\centering
\hspace{-1.cm}\includegraphics[width=2.2in,angle=0]{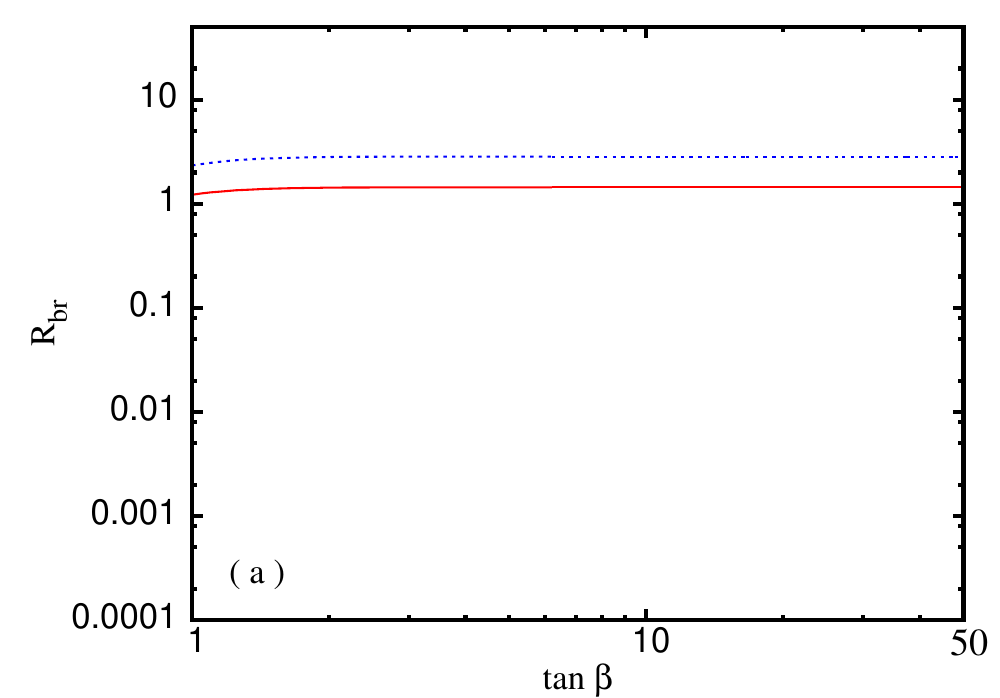}
\hspace{-.3cm}
\includegraphics[width=2.2in,angle=0]{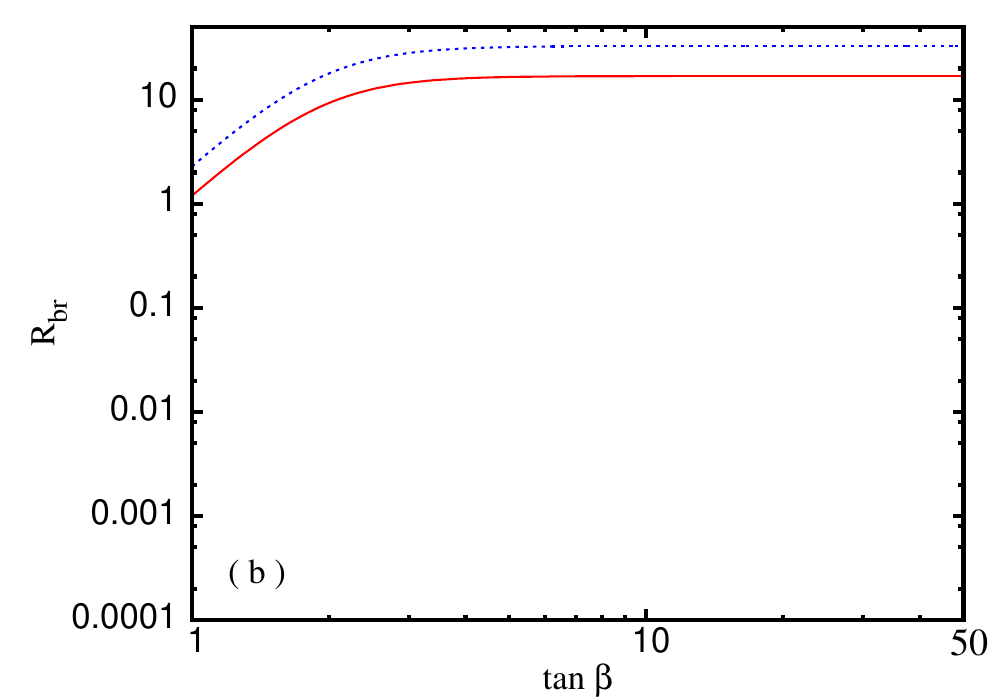}
\hspace{-.3cm}
\includegraphics[width=2.2in,angle=0]{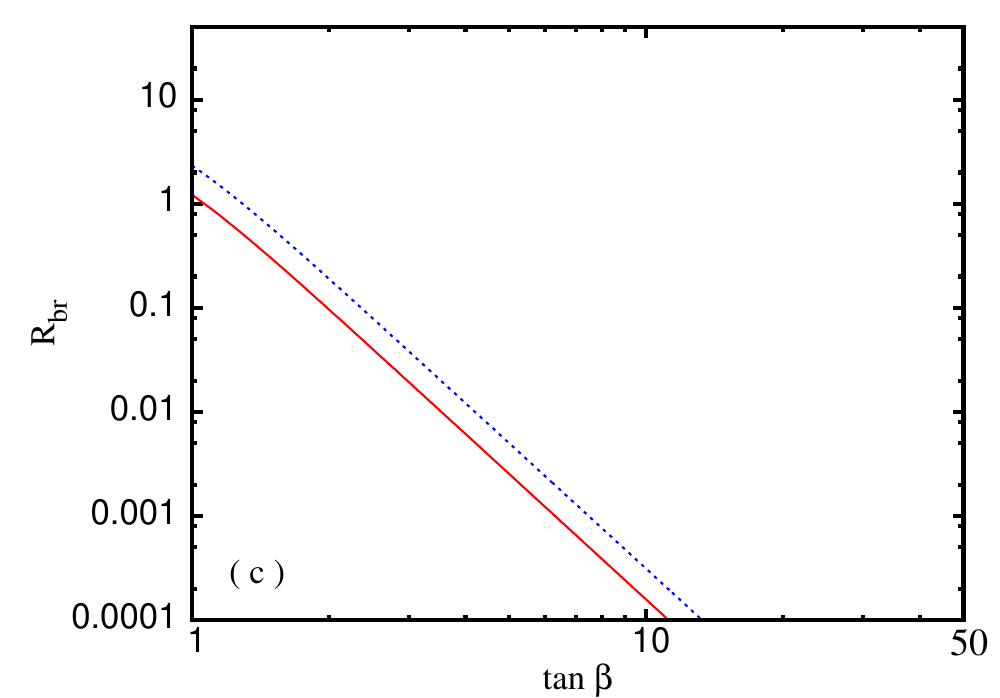}
\caption{The ratio $R_{br}$ as a function of $\tan\beta$ for $m_A=125$ GeV
  (solid line) and  $m_A=140$ GeV (dashed line) in 
2HDM-II,($a$) 2HDM-IV ($b$) and 2HDM-III ($c$). 
2HDM-I is not shown because the
$\tan\beta$ dependence drops in this ratio. }
\label{fig1}
\end{figure}

As one can see, in terms of branching ratios, we can have an enhancement
in $Br(A \to \tau \tau)$ in the 2HDM with respect to the SM, which can be
quantified by $R_{br}$,  in the following cases:
\begin{itemize}
\item 2HDM-II with large $\tan\beta$: $Br(A \to \tau \tau)^{II}\approx
  1.2 \times Br(H \to \tau \tau)^{SM}$,
\item 2HDM-IV with large $\tan\beta$: $Br(A \to \tau \tau)^{IV}\approx
  13 \times Br(H \to \tau \tau)^{SM}$.
\end{itemize}
It should be noted that values of $\tan \beta < 1$ are already excluded by the precision constraints in the 2HDM.
In Fig.~\ref{fig1}, we present the ratio $R_{br}$ as a function of $\tan\beta$
for $m_A=125$ GeV  (solid line) and  $m_A=140$ GeV (dashed line) 
in 2HDM-II ($a$), 2HDM-IV ($b$) and 2HDM-III ($c$). 
The 2HDM-I scenario is not shown because the
$\tan\beta$ dependence in this ratio cancels out.

Let us now turn to the ratio $R_\sigma$.  The $A$ and $h$ couplings
to $b\bar{b}$ are proportional to $\tan \beta$ for large $\tan\beta$
in Types II and III, as shown in Table~\ref{Tab:MixFactor}. 
Therefore, for large $\tan\beta$ only those
models can present an enhancement in the cross section coming
from the bottom loop in the gluon fusion process and from
$b\bar{b}$ fusion.
The only difference between $A$ and $h$ is in the 
mixing angle $\sin\alpha$ which is present in all
Yukawa couplings and will be discussed in the next section.
\begin{figure}[h!]
\centering
\hspace{-1.cm}\includegraphics[width=2.651in,angle=0]{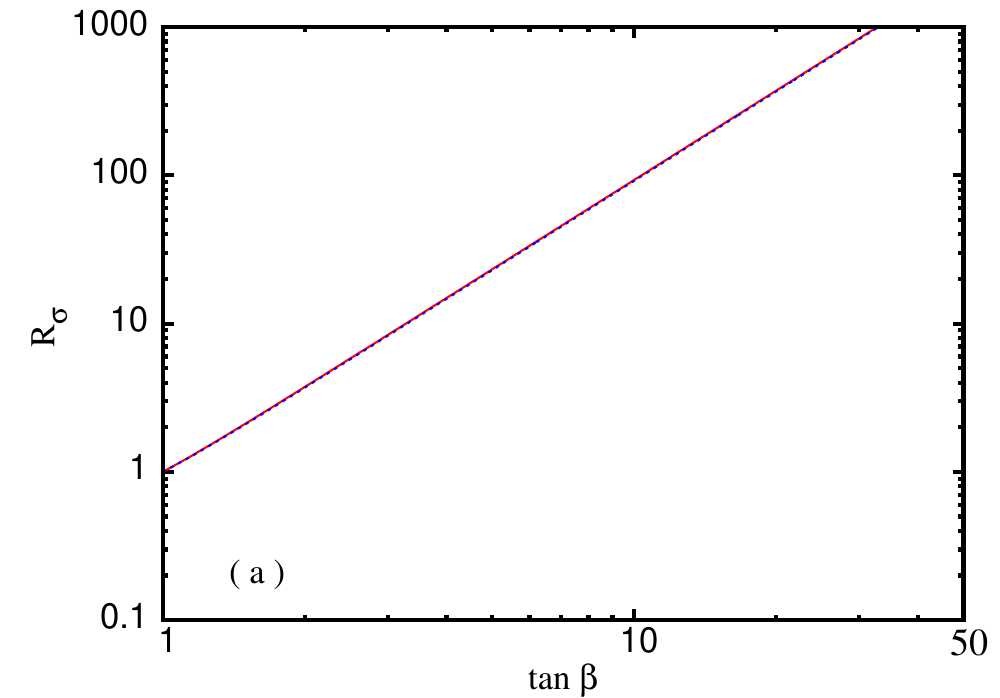}
\includegraphics[width=2.651in,angle=0]{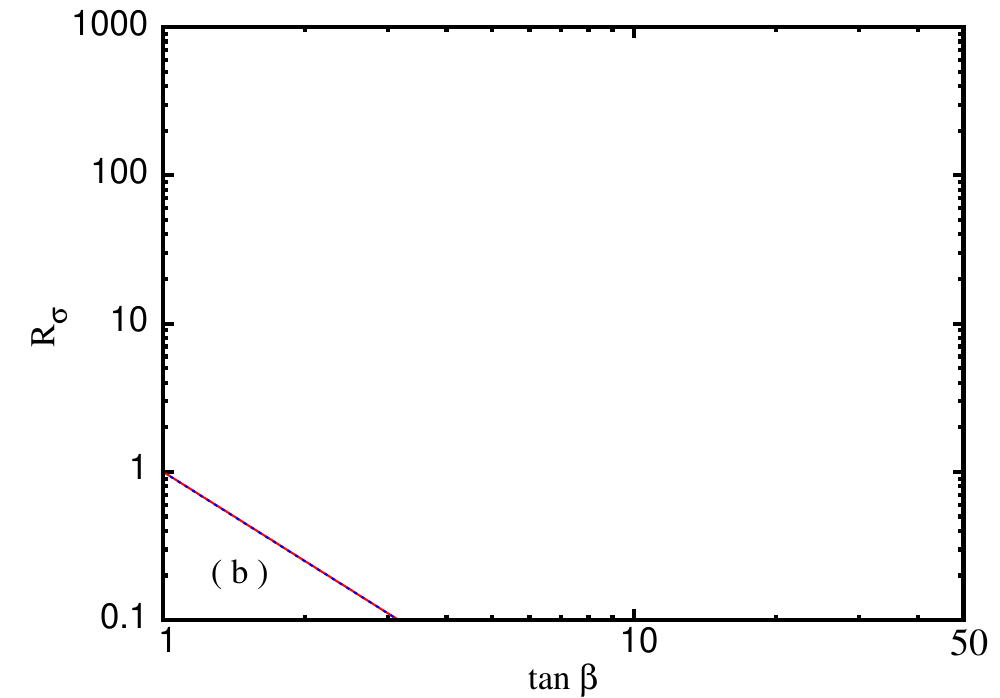}
\caption{The ratio $R_{\sigma}$ as a function of $\tan\beta$ 
for $m_A=125$ GeV (solid line) and  $m_A=140$ GeV (dashed line) in 
2HDM-II, III ($a$) and 2HDM-I,IV ($b$).}
\label{fig2}
\end{figure}
In Fig.~\ref{fig2} we present $R_{\sigma}$ as a function of $\tan\beta$ 
for $m_A=125$ GeV (solid line) and  $m_A=140$ GeV (dashed line) in 
2HDM-II, IV ($a$) and 2HDM-I,III ($b$). It is clear that an 
enhancement relative to the SM can only be obtained for large 
$\tan \beta$ and just for 2HDM-II and III.

%
%

\begin{figure}[ht]
\begin{tabular}{cc}
\hspace{-1.cm}\includegraphics[width=2.651in,angle=0]{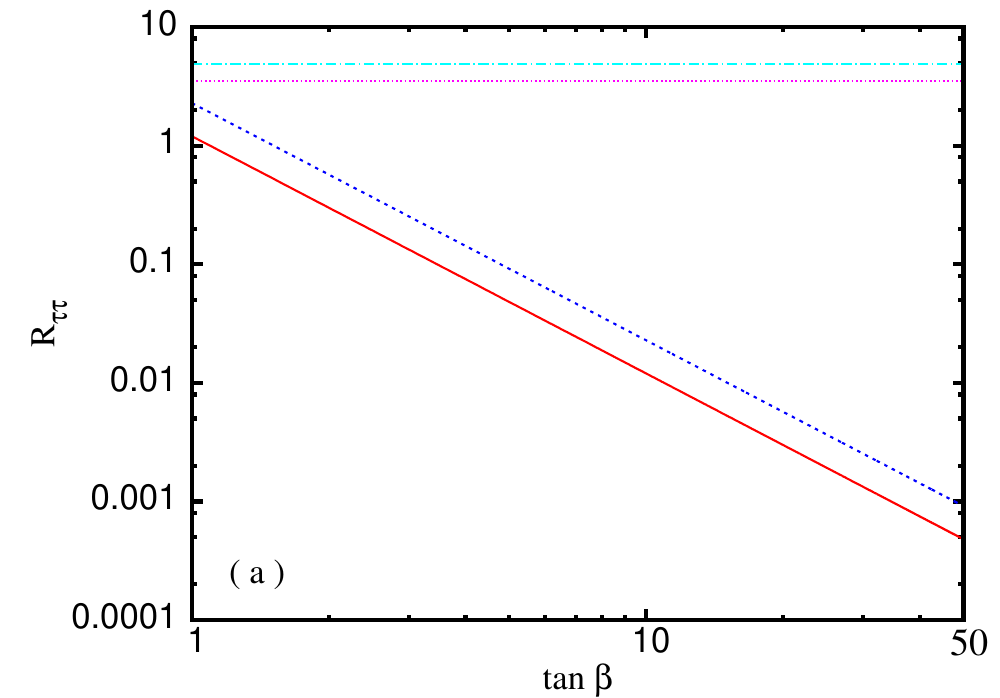}&
\includegraphics[width=2.651in,angle=0]{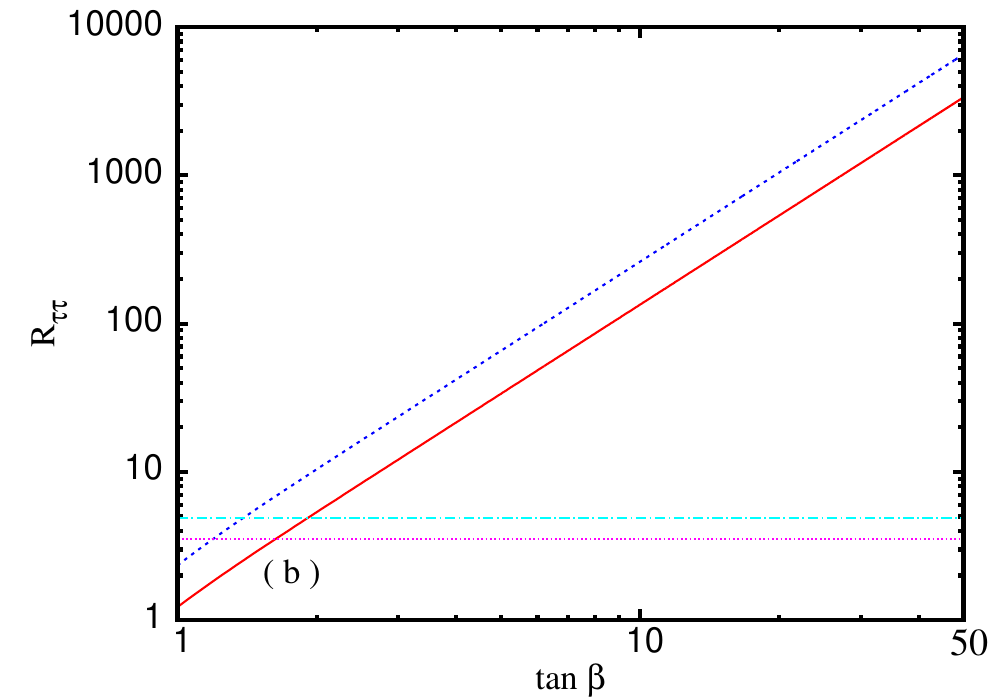}\\
\hspace{-1.cm}\includegraphics[width=2.651in,angle=0]{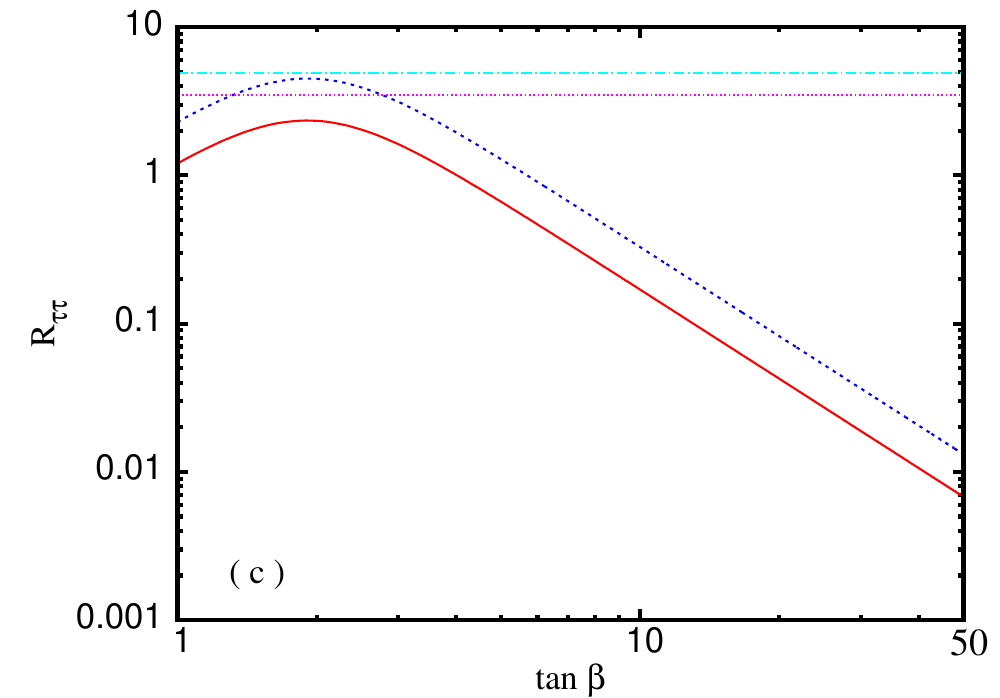}&
\includegraphics[width=2.651in,angle=0]{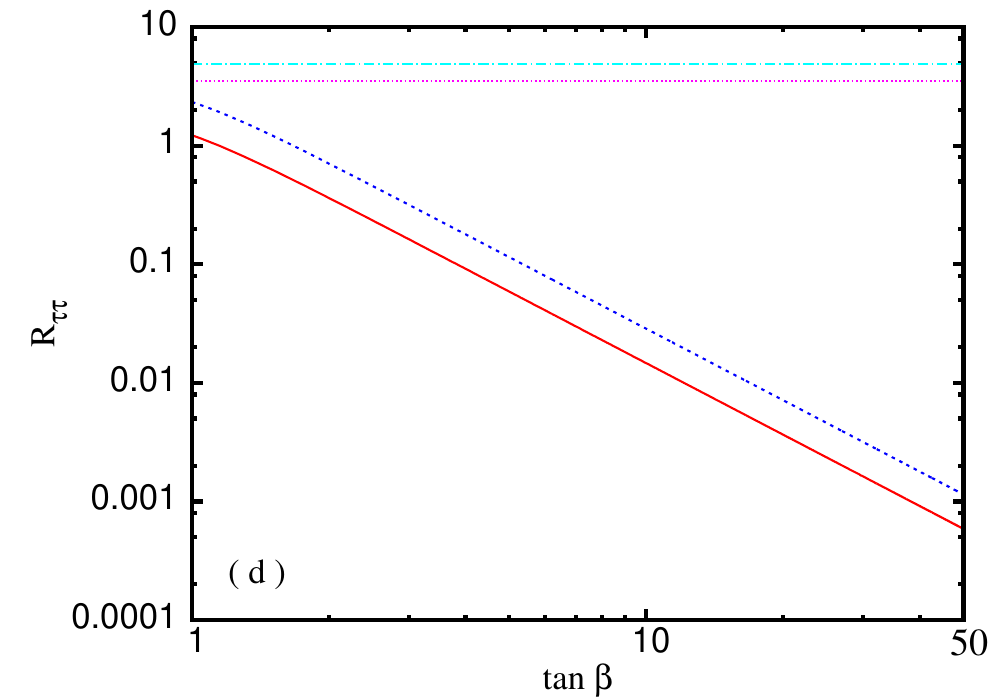}
\end{tabular}
\caption{The ratio $R_{\tau\tau}$ as a function of $\tan\beta$ 
for $m_A=125$ GeV
  (solid line) and  $m_A=140$ GeV (dashed line) in 
2HDM-I, ($a$), 2HDM-II ($b$),
2HDM-IV, ($c$), and 2HDM-III ($d$). The horizontal lines indicate
the $95\%$ CL CMS~\cite{CMS-PAS-HIG-11-029}  lower limit for $m_A=125$ GeV (lower line) 
and $m_A=140$  GeV (upper line). }
\label{fig3}
\end{figure}

It is now easy to understand the values of $R_{\tau\tau}$ presented in
Fig.~\ref{fig3} as a function of $\tan\beta$  for $m_A=125$ GeV
 (solid line) and  $m_A=140$ GeV (dashed line). The horizontal lines in 
the figures indicate the CMS $95\%$ CL lower limit  
for $m_A=125$ GeV (lower line) and $m_A=140$  GeV (upper line) 
\cite{CMS-PAS-HIG-11-029}. In 2HDM-II (Fig. \ref{fig3}$(b)$), 
for a given CP odd Higgs boson mass,
there is just a small window of $\tan \beta $ which is allowed by the CMS data on 
$R_{\tau \tau}$. It is clear that, for 2HDM-II, 
values of $\tan\beta\geq 2$ are definitely excluded at $95\%$ CL 
for the entire mass range under scrutiny. In 2HDM-IV, the excluded
region reduces to a tiny window centred at $\tan\beta=2$, while 
for 2HDM-I and III $\tan\beta$ is allowed in the entire range shown. 

\subsection{CP-even case}

We will now move on to the the study of the lightest CP-even scalar. 
Although similar, the process $pp (gg + b\bar{b}) \to h \to \tau^+ \tau^-$, 
is in fact more evolved since the Yukawa couplings of $h$ depend both
on $\tan\beta$ and $\alpha$. 
Throughout this section we take $\sin \alpha > 0$
to make the plots clearer. There is an approximate
symmetry in the values of $R_{\tau \tau}$ 
between positive and negative values of $\sin \alpha$. In the final exclusion plots, 
we will allow $\sin \alpha$ to vary in the entire range from $-1$ to $1$ though.
We shall now discuss each Yukawa model in turn.

\subsubsection{2HDM-I}
\begin{figure}[hI]
\centering
\hspace{-1.cm}\includegraphics[width=2.21in,angle=0]{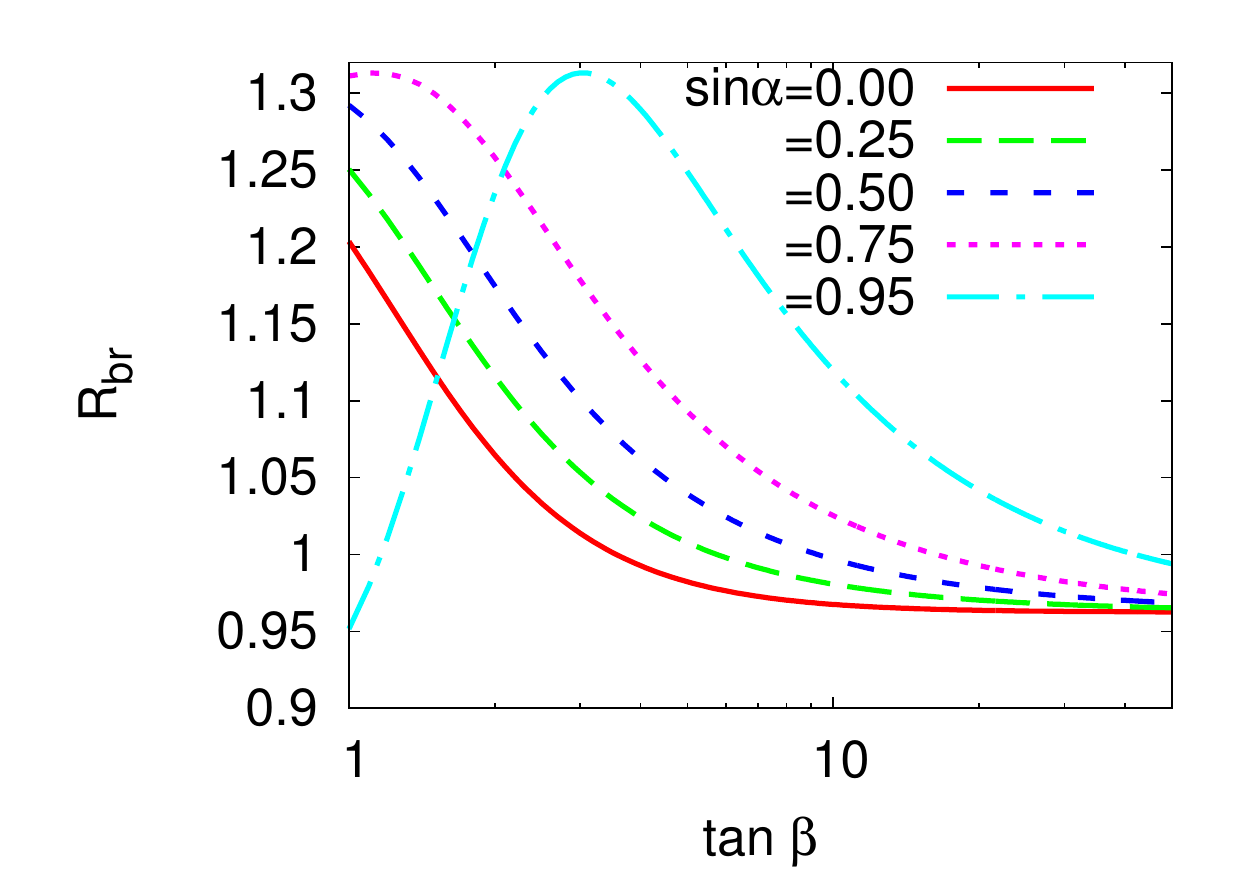}
\hspace{-.3cm}
\includegraphics[width=2.21in,angle=0]{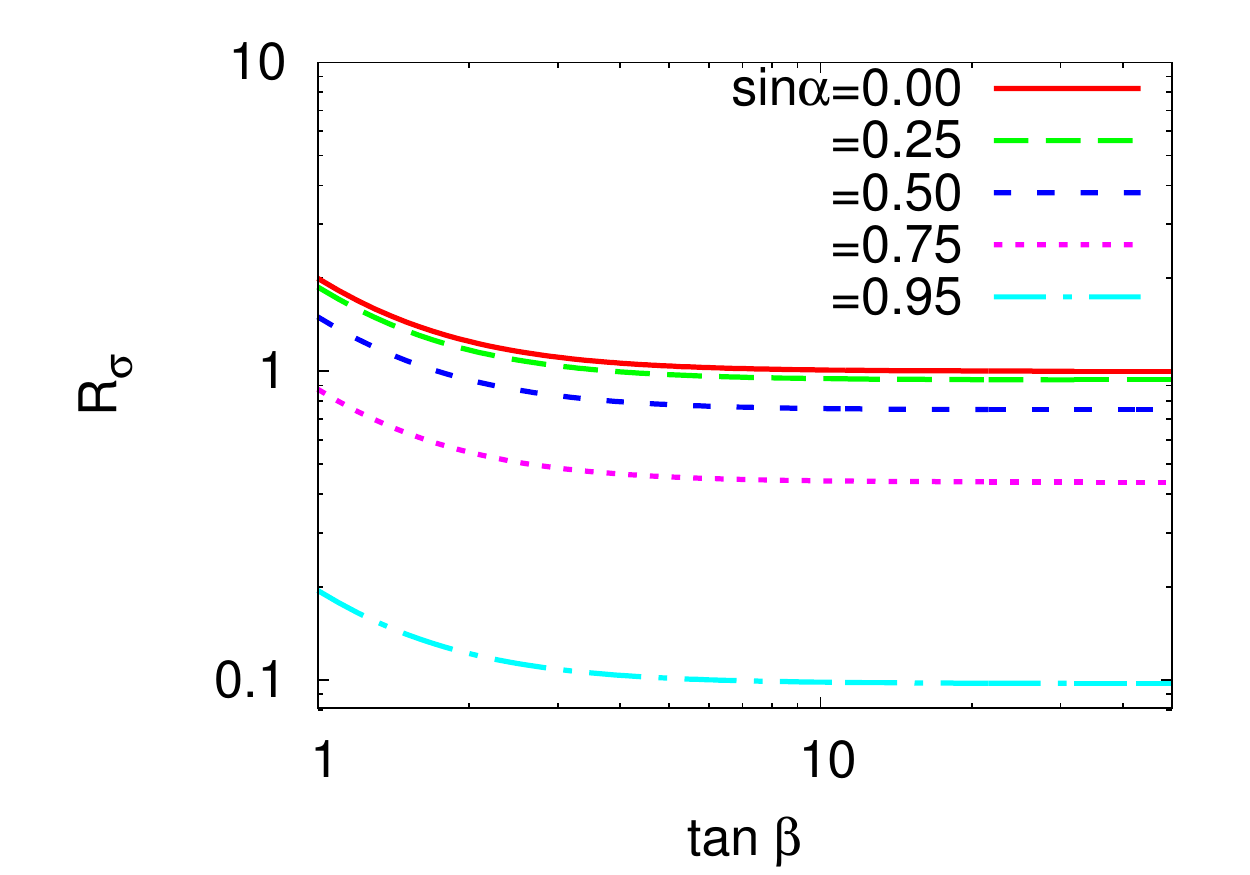}
\hspace{-.3cm}
\includegraphics[width=2.21in,angle=0]{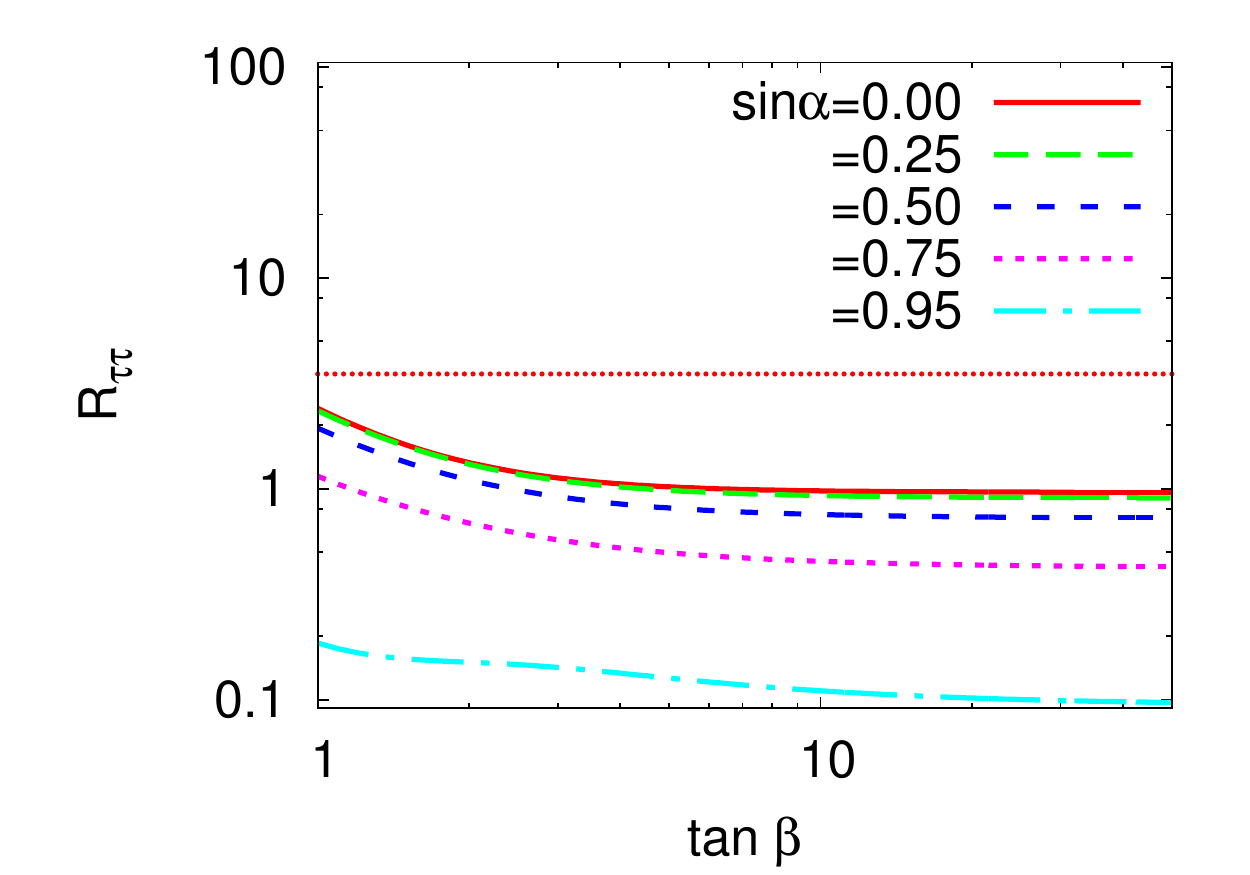}
\caption{$R_{br}^{h}$, $R_\sigma$ and $R_{\tau\tau}$ as functions of $\tan\beta$ for several values of $\sin\alpha$ with $m_h=125$ GeV in 2HDM-I. 
The horizontal dotted line indicates the $95\%$ CL CMS lower limit 
on $R_{\tau \tau}$ for $m_h=125$ GeV.}
\label{fig4}
\end{figure}

In 2HDM-I, all fermions couple to $h$ proportionally to
$y_h^\tau=y_h^b=y_h^c$=$\cos\alpha\over\sin\beta$. 
Such a Higgs would be fermiophobic in the limit
$\cos\alpha=0$ ($\sin\alpha=\pm 1$) where the production cross section
would be reduced. Since $\tan \beta > 1$, $R_\sigma$ is
approximately constant in all $\tan \beta$ range   
as shown in Fig.~\ref{fig4} (middle) where $R_\sigma$ is shown
as a function of $\tan\beta$ for several values of $\sin\alpha$ 
and $m_h=125$ GeV.

The branching ratio of the CP-even Higgs boson, $h$, 
to the $\tau\tau$ mode takes the following form
\begin{eqnarray}
Br(h\to \tau \tau)&=&\frac{m_\tau^2 (y_h^\tau)^2}{m_\tau^2 (y_h^\tau)^2+ 3
  m_b^2 (y_h^b)^2+ 
3 m_c^2 (y_h^c)^2+ F (\Gamma_{WW^*}, \Gamma_{ZZ^*}) } ~,
\label{brhtau}
\end{eqnarray}
where $F (\Gamma_{WW^*}, \Gamma_{ZZ^*})$ is proportional 
to the partial widths of $h\to WW^*$ and
$h\to ZZ^*$, and the former is about 0.2 $\Gamma_{Total}^{SM}$ 
for a Higgs boson of 125 GeV.
Since $y_h^\tau = y_h^b = y_h^c$ in 2HDM-I,
this common factor will cancel out in the branching ratio 
formula Eq.~(\ref{brhtau}), rendering a $Br(h \to \tau \tau)^I$ similar to the
corresponding SM rate. 
This is illustrated in Fig.~\ref{fig4} (left), where we can see
that the ratio $R_{br}$ is close to one except for some modulations
 due to the $h\to WW^*$ contribution, which is
proportional to $\sin(\beta-\alpha)$.
At large $\tan\beta$, the contribution of the vector
bosons decays lose their angular dependence,
amounting to a reduction in $Br(h \to \tau \tau)^I$ as shown 
in Fig.~\ref{fig4} (left).

From the above discussions, one can readily understand the shape of 
$R_{\tau\tau}$ in 2HDM-I presented in Fig.~\ref{fig4} (right). 
$R_{\tau\tau}$ is almost independent of $\tan \beta$ in the
entire range shown and decreases with increasing $\sin \alpha$
due to the cross section behaviour. It is also clear from the plots that 
there are no excluded regions in the 2HDM-I parameter space
from the recent LHC data on $R_{\tau\tau}$.

\subsubsection{2HDM-II}
In 2HDM-II, as seen in Table~\ref{Tab:MixFactor}, the $h$ couplings to a pair of $\tau$ leptons and to $b \bar b$ are proportional to $y_h^\tau=y_h^b$=$\sin\alpha\over\cos\beta$.
These couplings will boost the $b\bar{b}\to h$ cross section
at high $\tan\beta$ as compared to the SM one. The $h$ coupling 
to top-quarks is proportional to $y_h^u$=$\cos\alpha\over\sin\beta$.  This implies that the top loop in the $gg\to h$ cross section could only be enhanced at small 
$\tan\beta$, while for large $\tan\beta$ the bottom loop 
will take over.

\begin{figure}[ht]
\centering
\hspace{-1.cm}\includegraphics[width=2.21in,angle=0]{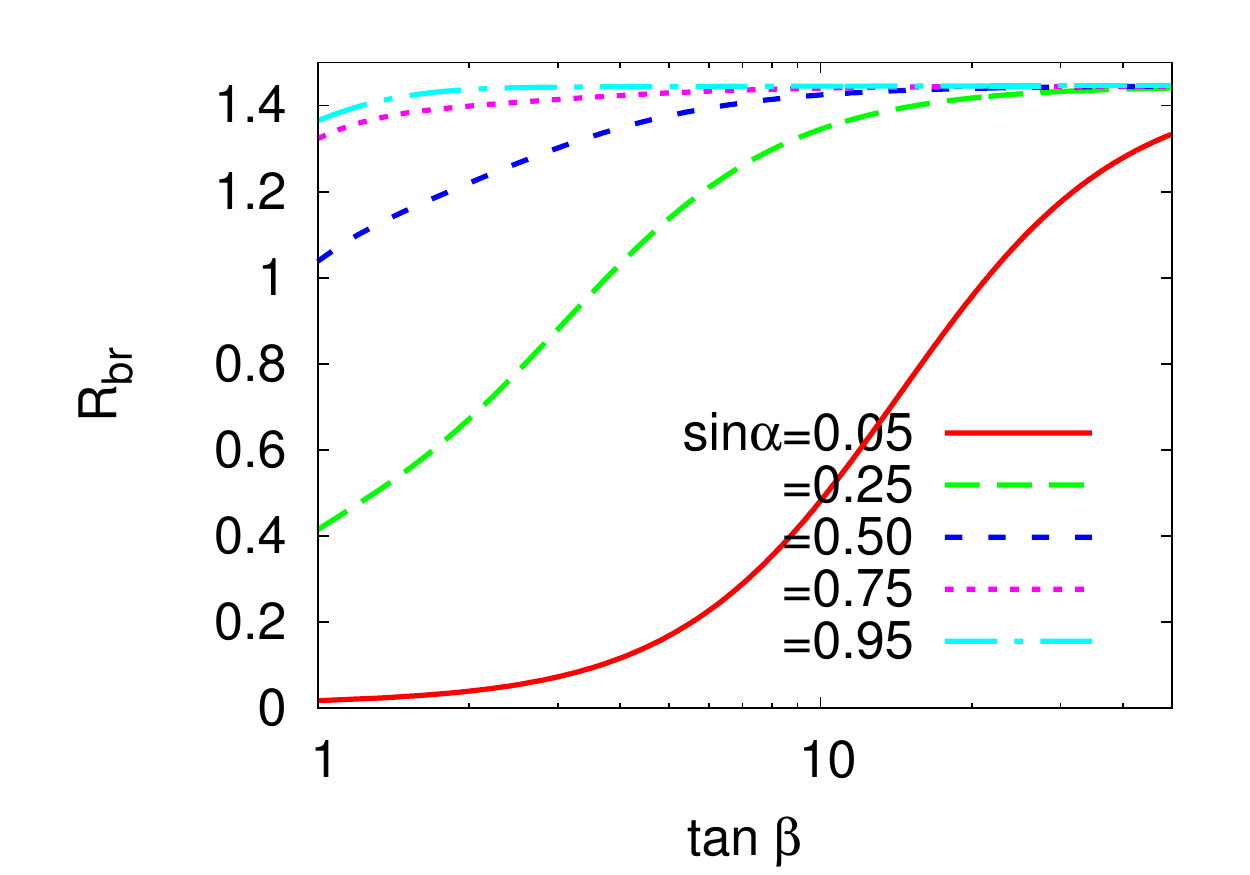}
\hspace{-.3cm}
\includegraphics[width=2.21in,angle=0]{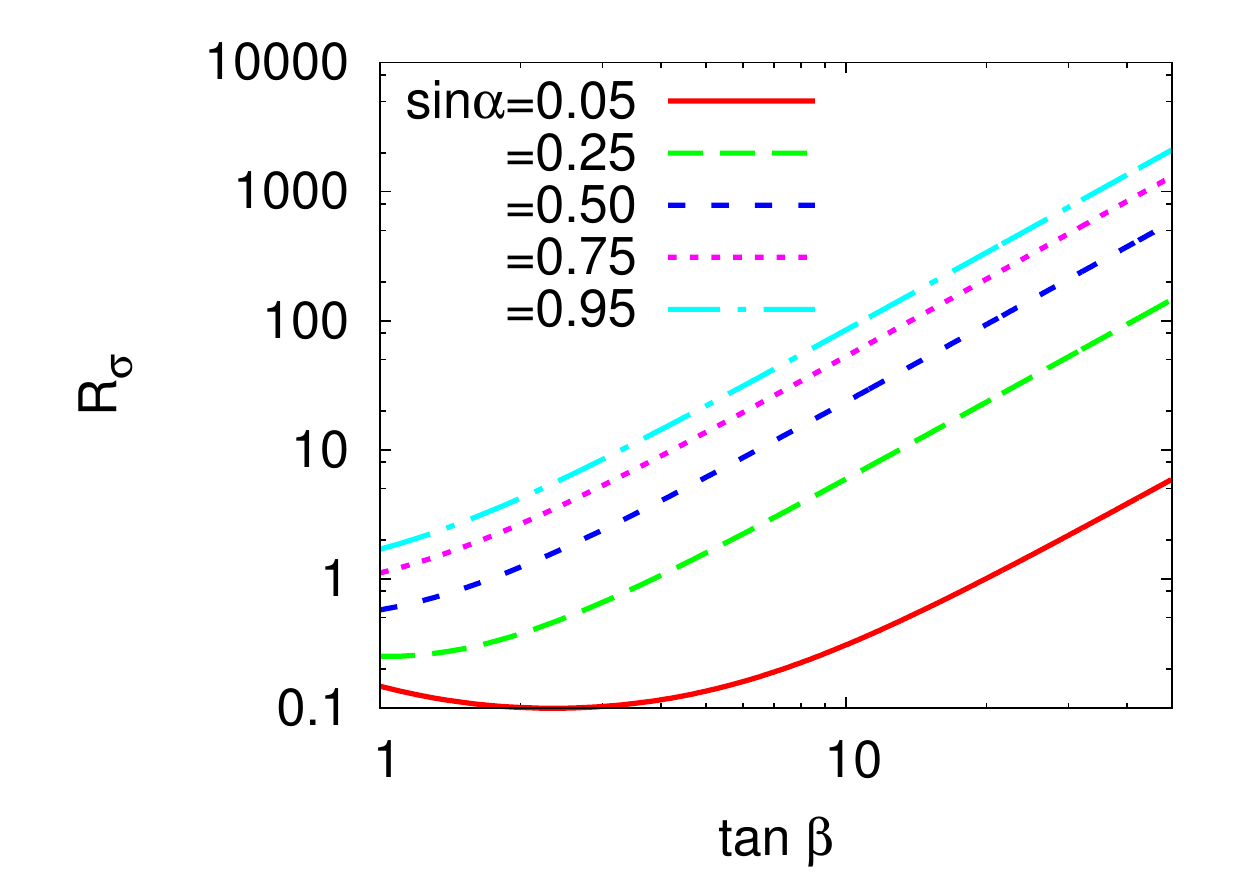}
\hspace{-.3cm}
\includegraphics[width=2.21in,angle=0]{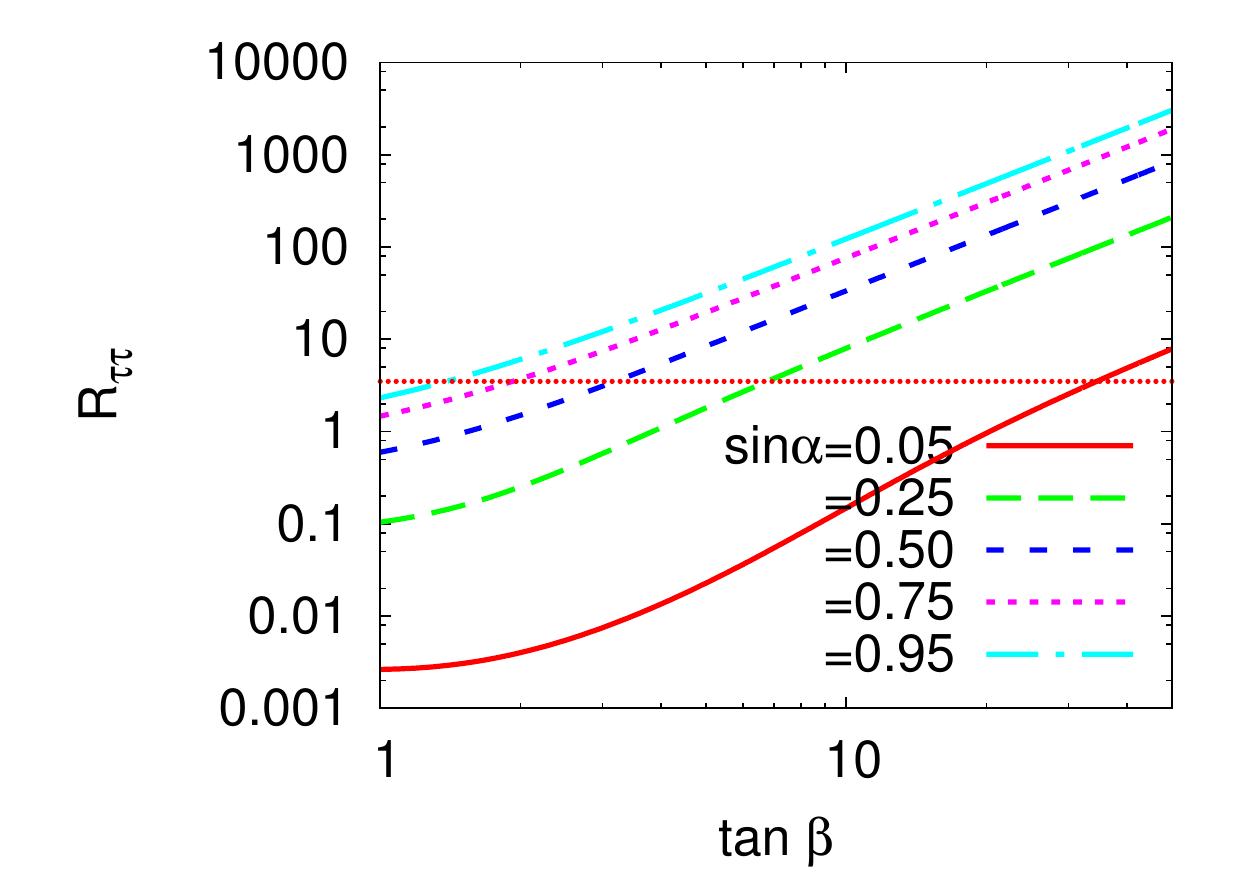}
\caption{$R_{br}^{h}$, $R_\sigma$ and $R_{\tau\tau}$ as a function of
  $\tan\beta$ for several values of $\sin\alpha$ 
with $m_h=125$ GeV in the  2HDM-II. 
The horizontal dotted line indicates the $95\%$ CL CMS lower limit 
on $R_{\tau \tau}$ for $m_h=125$ GeV.}
\label{fig5}
\end{figure}

In 2HDM-II, a large $\tan\beta$ does
modify significantly the branching ratio of $h \to \tau^+\tau^-$
relative to the SM one. The only possible enhancement factor,  
$1/\cos\beta$ [Eq.~(\ref{brhtau})] will cancel amongst
the main fermion contributions.
Therefore, $R_{br}\approx 1.4$ for large $\tan\beta$ as 
can be read from Fig.~\ref{fig5} (left). 
When $\tan\beta$ is small, the branching ratio of $h \to \tau^+\tau^-$
is suppressed.  This suppression is even more pronounced for smaller
$\sin\alpha$ and when $\sin\alpha\to 0$,  $Br(h \to \tau^+\tau^-)\to 0$ and
consequently $R_{br}\to 0$.
 
On the other hand, the production cross section via both gluon and 
$b\bar{b}$ fusion is enhanced with respect to the SM Higgs production rate
for large $\tan\beta$ and not too small $\sin\alpha$, as can be 
seen in Fig.~\ref{fig5} (middle). 

Finally, the ratio $R_{\tau\tau}^{II}$ is shown in Fig.~\ref{fig5} (right). 
Taking into account the dotted line of the $95\%$ CL CMS lower 
limit on $R_{\tau \tau}$ for $m_h=125$ GeV
we conclude that a substantial part of the $\tan \beta$ range presented
is excluded, except for very small values of $\sin \alpha$.

\subsubsection{2HDM-III}
\begin{figure}[ht]
\centering
\hspace{-1.cm}\includegraphics[width=2.21in,angle=0]{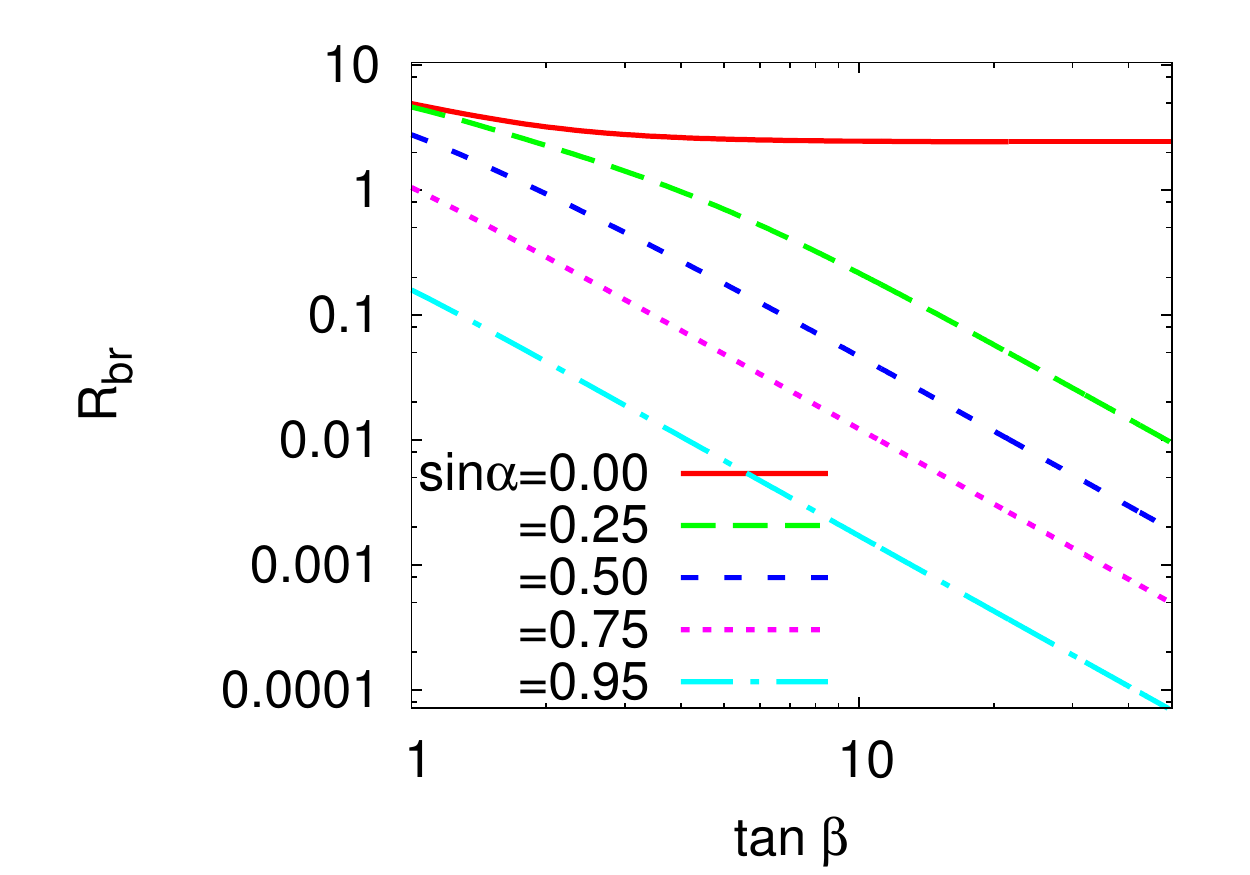}
\hspace{-.3cm}
\includegraphics[width=2.21in,angle=0]{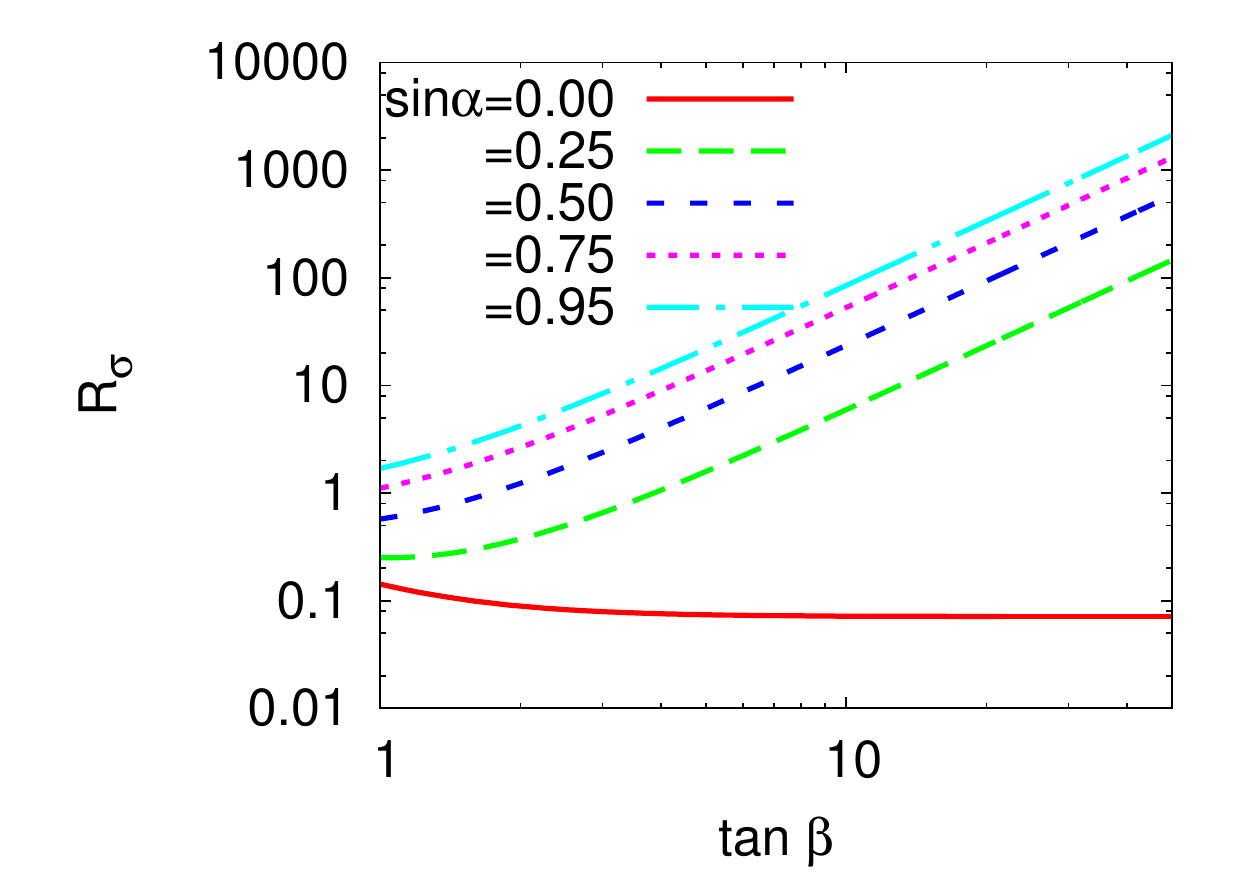}
\hspace{-.3cm}
\includegraphics[width=2.21in,angle=0]{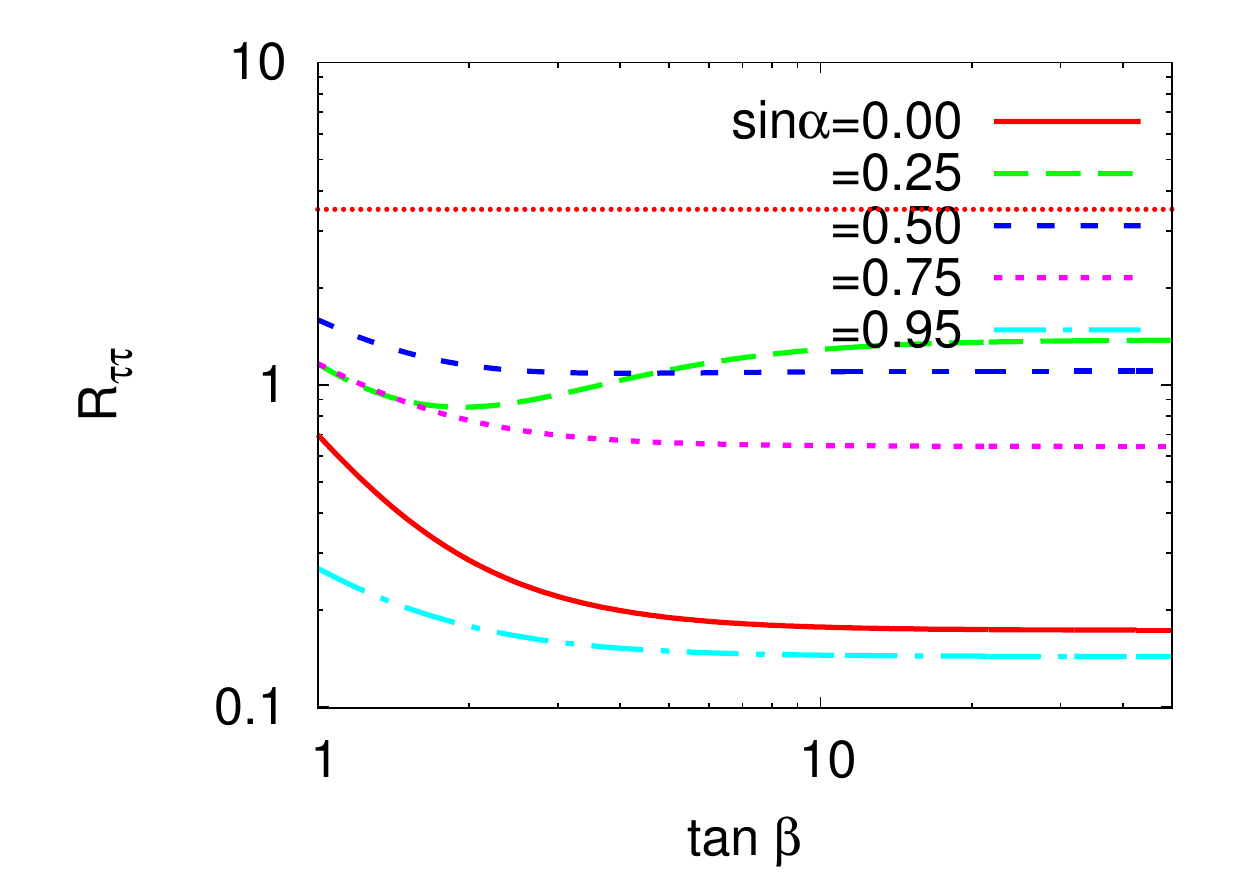}
\caption{$R_{br}^{h}$, $R_\sigma$ and $R_{\tau\tau}$ as functions of 
$\tan\beta$ for several values of $\sin\alpha$ with $m_h=125$ GeV in 2HDM-III. 
The horizontal dotted line indicates the $95\%$ CL CMS lower limit 
on $R_{\tau \tau}$ for $m_h=125$ GeV.}
\label{fig6}
\end{figure}

In 2HDM-III, the $h$ couplings to a pair of fermions are proportional to
$y_h^\tau=y_h^c$=$\cos\alpha\over\sin\beta$, 
$y_h^b$=$-\sin\alpha\over\cos\beta$.
Therefore, as in model II, the bottom Yukawa coupling will be enhanced for 
large $\tan\beta$.  As $\tan\beta$ increases, the partial
width of the $h\to b\bar{b}$ becomes large and thus suppresses 
$Br(h \to \tau^+\tau^-)$. This is illustrated in Fig.~\ref{fig6} (left), 
where the role played by $\sin\alpha$ is clearly observed.
Of course for $\sin\alpha=0$, $h\to b\bar{b}$ vanishes and 
$Br(h \to \tau^+\tau^-)$ is at least more than 2 times larger than 
the SM rate for almost all values of $\tan\beta$.

In Fig.~\ref{fig6} (middle), we see the enhancement of the cross section at large $\tan \beta$
due to the b-quark loop and the $b \bar b$ production cross section. 
However, once combined with $R_{br}$, the ratio 
$R_{\tau\tau}$ is always well below he CMS lower limit at 
$95\%$ CL. Therefore the entire range of $\tan \beta$
is allowed, independently of the values of $\sin \alpha$.

\subsubsection{2HDM-IV}
\begin{figure}[ht]
\centering
\hspace{-1.cm}\includegraphics[width=2.21in,angle=0]{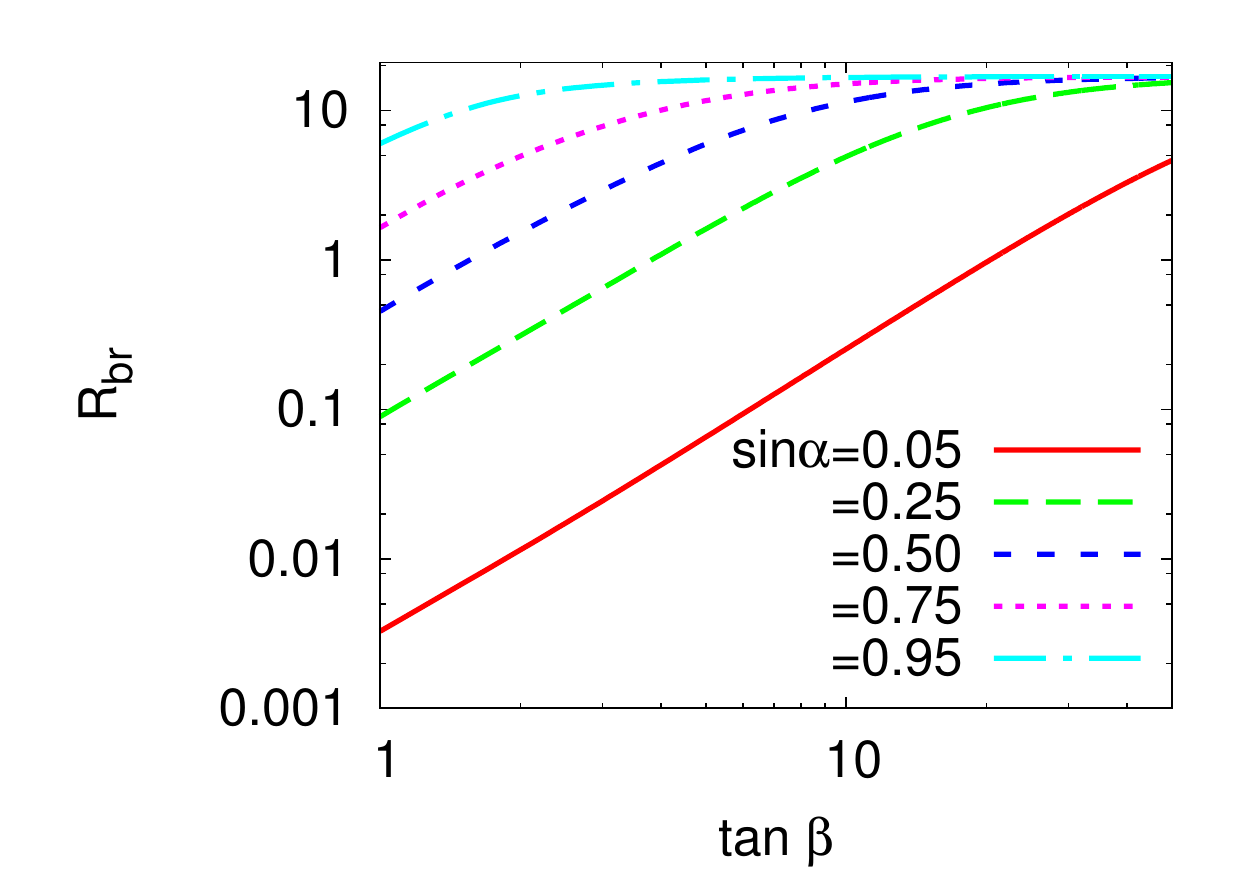}
\hspace{-.3cm}
\includegraphics[width=2.21in,angle=0]{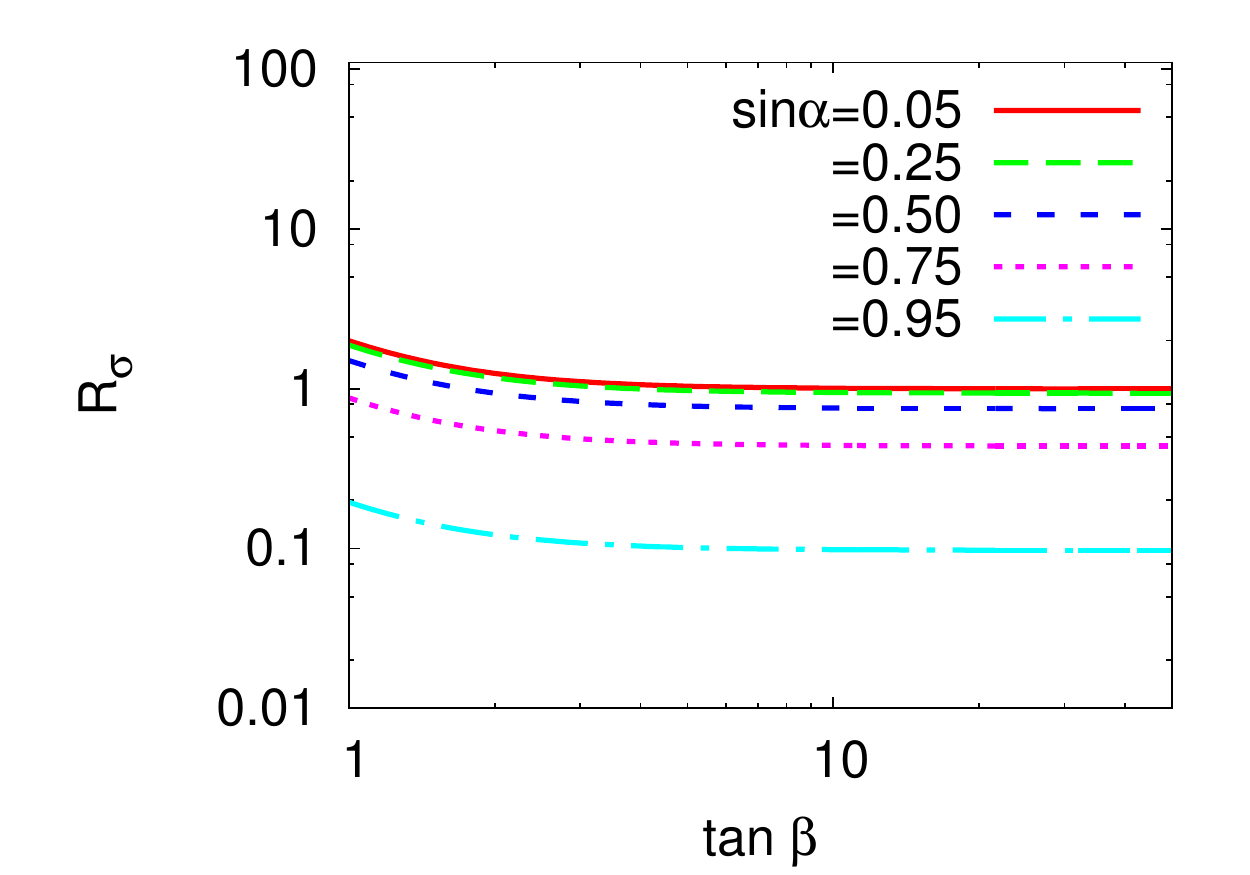}
\hspace{-.3cm}
\includegraphics[width=2.21in,angle=0]{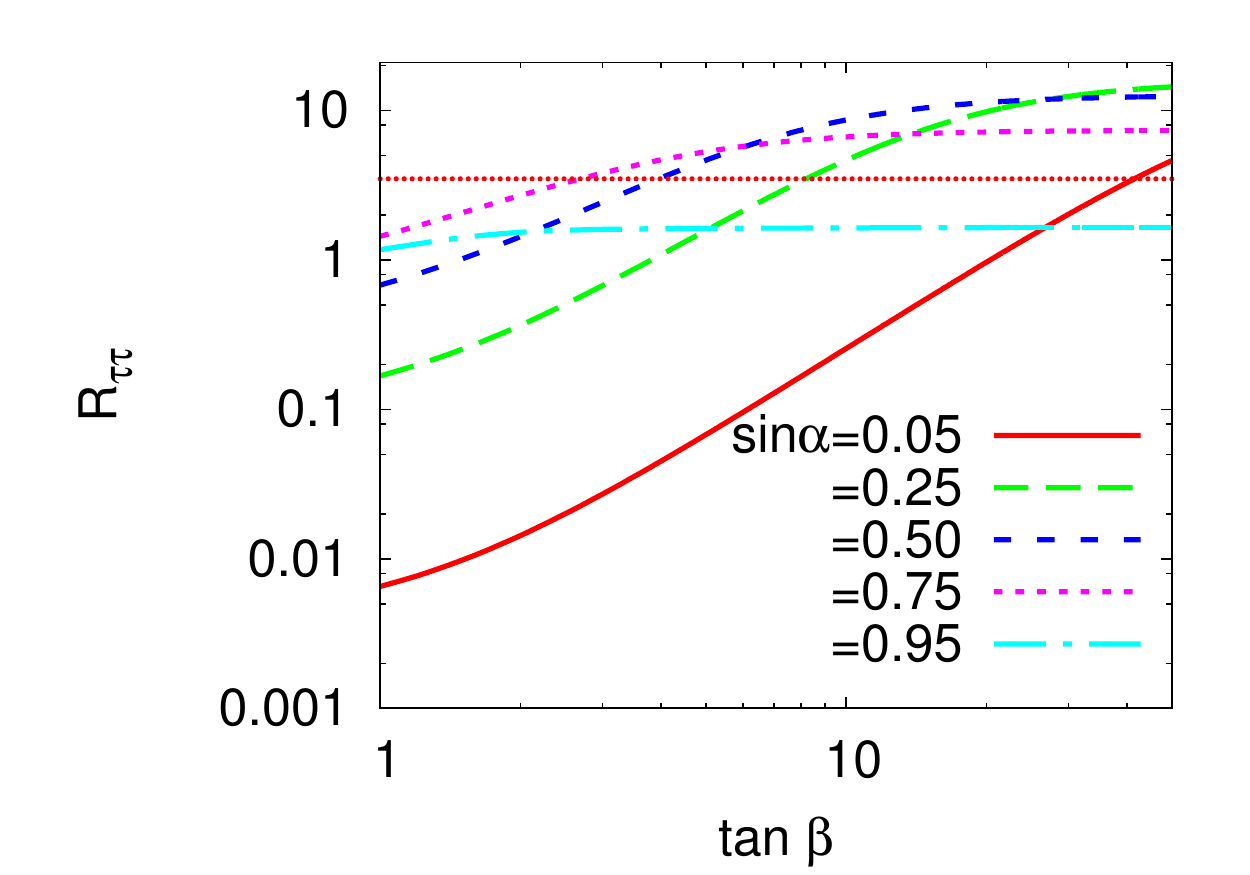}
\caption{$R_{br}^{h}$, $R_\sigma$ and $R_{\tau\tau}$ as functions of 
$\tan\beta$ for several values of $\sin\alpha$ with $m_h=125$ GeV in 
2HDM-IV. The horizontal dotted line indicates the CMS limit for $m_h=125$ GeV.}
\label{fig7}
\end{figure}

In 2HDM-IV (also known as the lepton-specific model), the $h$ couplings 
to a pair of fermions are $y_h^\tau$=$-\sin\alpha\over\cos\beta$ and 
$y_h^b=y_h^c$= $\cos\alpha\over\sin\beta$, relative to the SM ones.
Hence, only $y_h^\tau$ can be enhanced relative to the SM at 
large $\tan\beta$. As stated previously, in 2HDM-IV the 
$Br(h \to \tau^+\tau^-)$ can be $100\%$ at large $\tan\beta$ and not too
small $\sin\alpha$. This can be seen in Fig.~\ref{fig7} (left), 
where $R_{br}$ can reach values slightly larger than 10. 

As for model I, the production cross section relative to the SM 
are almost independent of $\tan \beta$ for the entire range presented in 
in Fig.~\ref{fig6} (middle). There is however a $\sin \alpha$ dependence:
$R_\sigma$ decreases with increasing $\sin \alpha$.  
When $\sin\alpha\to 1$, it is clear that the production cross section will vanish since $y_h^b=y_h^c$ is proportional to $\cos\alpha\to 0$.

The ratio $R_{\tau\tau}^{IV}$ is shown in Fig.~\ref{fig6} (right).  
At large $\tan\beta$, we can see the effect
of the $R_{br}$ enhancement.  The excluded region depends
heavily on $\sin \alpha$ and will be presented in the conclusions.

\section{A very light CP-odd boson \label{sec:light}}
A very light CP-odd Higgs boson is not excluded in the proposed 
2HDM~\footnote{Some attentions have been recently given to the very 
low CP-odd Higgs boson mass region, decaying into muons, in the 
next-to-Minimal Supersymmetric Model~\cite{Dermisek:2009fd,Almarashi:2011hj}}. 
In fact, the LEP bounds on the 2HDM parameters based on the 
$e^+ e^- \to h A $~\cite{Schael:2006cr} production process, 
do not hold if either $\sin (\beta -\alpha) \approx 1$ or if the
mass of the lightest Higgs is such that the production
is either disallowed or its rate is very small.  However, LEP 
has also produced bounds that depend solely on the Yukawa 
couplings of the models.

There were searches at LEP based on the Yukawa processes 
$e^+ e^- \to b \bar{b} A  \to b \bar{b} \tau^ + \tau^ -$ in 
the mass range $m_A = 4 - 12$ GeV ~\cite{Abbiendi:2001kp} and in
the channels $b \bar{b}   b \bar{b}$, $b \bar{b}    \tau^ + \tau^ -$  
$ \tau^ + \tau^ - \tau^ + \tau^ -$
for pseudoscalar masses up to 50 GeV ~\cite{Abdallah:2004wy}. These
bounds on the Yukawa couplings can be divided into three groups: bounds
on $g_{A \tau^+ \tau^-}^2$, $g_{A \bar{b} b}^2$ and 
$g_{A \bar{b} b}  \, g_{A \tau^+ \tau^-}$.
In 2HDM-I, we have $g_{A \tau^+ \tau^-}^2 = g_{A \bar{b} b}^2=g_{A \bar{b} b}  \, g_{A \tau^+ \tau^-} = 1/ \tan^2 \beta$ and,
consequently, a lower bound on $\tan \beta$ can be extracted. In model type II, 
the relation is $g_{A \tau^+ \tau^-}^2 = g_{A \bar{b} b}^2=g_{A \bar{b} b}  \, g_{A \tau^+ \tau^-} = \tan^2 \beta$
and, therefore, an upper bound on $\tan \beta$ can be obtained as a function of the CP-odd scalar mass. Finally in models type III and IV,
no bounds on $\tan \beta$ can be derived in the mass region $m_A = 4 - 12$ GeV because
$g_{A \bar{b} b}  \, g_{A \tau^+ \tau^-} = 1$.

Recently, the ATLAS Collaboration has searched for a very light 
CP-odd scalar~\cite{ATLASmumu}
in the mass region $4-12$ GeV with the process $pp \to A \to \mu^+ \mu^-$. 
As previously
discussed, the cross section for the production of a CP-odd particle can 
only increase 
for large $\tan \beta$ in models type II and III --- the top 
quark couples to the CP-odd
scalar proportionally to $1/\tan \beta$, while the bottom quark 
couples proportionally to 
$\tan \beta$ in those scenarios.  On the other hand, the 
largest possible value for
$Br(A \to \mu^+ \mu^-)$ arises in Type IV while in Type III the branching
ratio to fermions is negligible for large $\tan \beta$. 
Therefore, in the high 
$\tan \beta$ region, one would expect
to reach the highest possible values for $\sigma(pp \to A \to \mu^+ \mu^-)$ 
in Types II and IV. Note that when one moves to values of $\tan \beta$ below 1, 
the top-quark
loop dominates in the production process that is similar in all scenarios.  As
the ATLAS result pertains only to the mass region $4-12$ GeV, we can only
compare their result with those obtained by the OPAL 
Collaboration~\cite{Abbiendi:2001kp}. 

\begin{table}[h!]
\begin{center}
\begin{tabular}{|c|c|c|c|c|c|c|c|c|c|c|c|}
\hline
& $m_A$ (GeV)          &  4 &  5 &  6 & 7 & 8 & 9 & 10 & 11 & 12 \\ \hline
& $\tan \beta_{Br(A \to \tau^+ \tau^-) = 100 \%}^{II}$  &  8.5  &  11.0 & 9.6 & 11.5 & 10.7 & 11.0 & 11.3 & 12.3 & 13.6\\ \hline
& $\tan \beta^{II}$  &  $\approx$ 8.5  &  $\approx$ 11.0 & $\approx$ 9.6 & $\approx$ 11.5 & $\approx$ 10.7 & $> 100$  & $> 100$ & $>100$ & $>100$\\
\hline
\end{tabular}
\end{center}
\caption{Bounds from the OPAL Collaboration on $\tan \beta$ for Type II with $m_A$ varying between 4 to 12 GeV. The second line shows the limits assuming 
$Br(A \to \tau^+ \tau^-) = 100\% $. The last line in the table takes into account the actual values of $Br(A \to \tau^+ \tau^-)$ in Type II.}
\label{Tab:OPALdata}
\end{table}

The OPAL results are shown in Table~\ref{Tab:OPALdata} in the form of a limit
on $\tan \beta$ as a function of the scalar mass.  In the second line of the table 
the limits assume $Br(A \to \tau^+ \tau^-) = 100 \%$, while in the last line 
the actual values of $Br(A \to \tau^+ \tau^-)$ for Type II are taken
into account.  Using the ATLAS result on $pp \to A \to \mu^+ \mu^-$, we can 
now inquire whether the bounds on $\tan \beta$ for $m_A = 4 - 12$ GeV are improved
relative to the OPAL bounds. In our calculation, we do not take into account
the appearance of bound states.

\begin{figure}[h!]
\centering
\hspace{-1.cm}
\includegraphics[width=2.8in,angle=0]{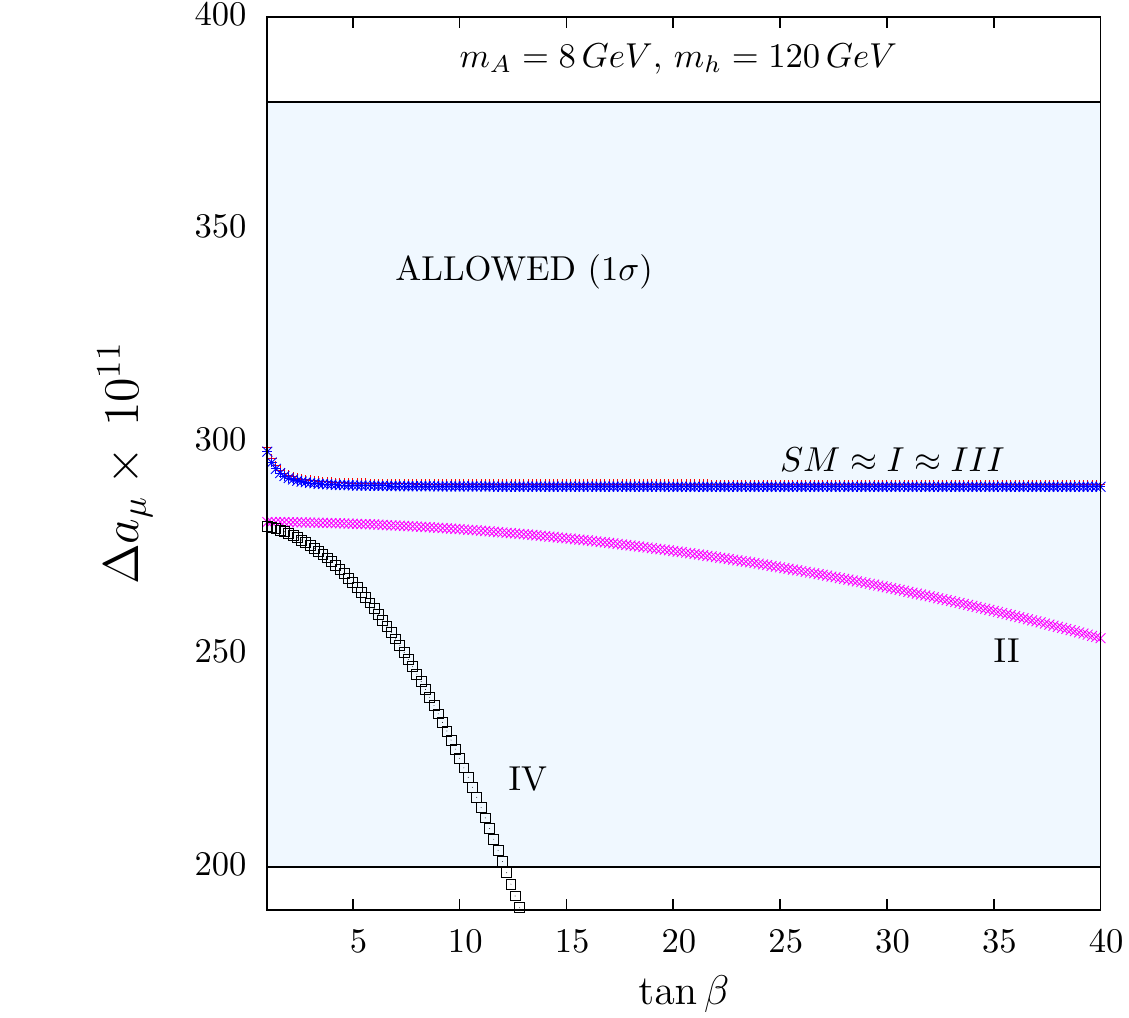}
\includegraphics[width=2.8in,angle=0]{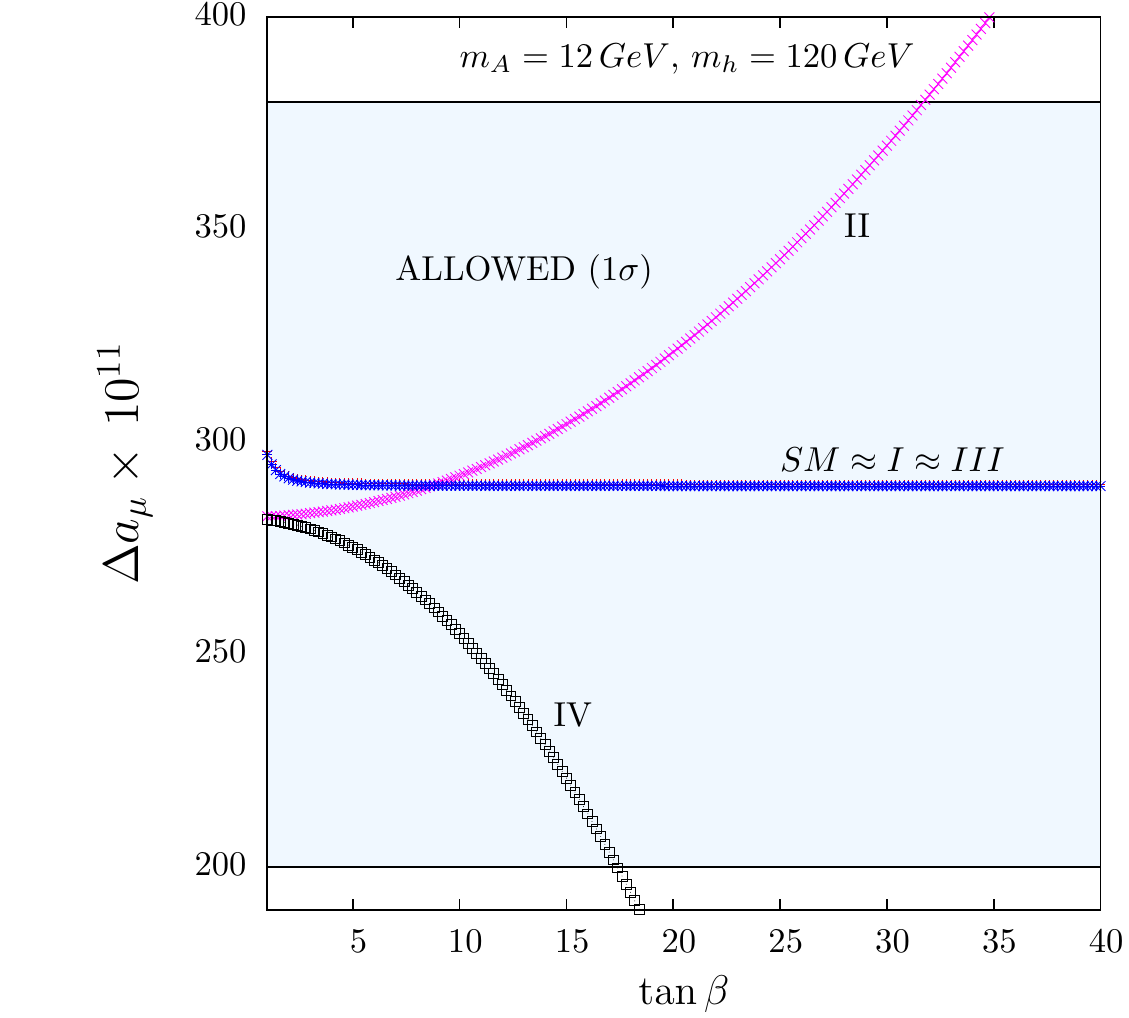}
\caption{The left panel shows $\Delta a_{\mu} = (a_{\mu}^{exp} - a_{\mu}^{th-SM} ) +  a_{\mu}^{th-2HDM}$ as a function of $\tan \beta$ for $m_A = 8$ GeV and $m_h = 120$ GeV in the limit $\alpha \approx \beta$.  The same is for the right panel except $m_A = 12$ GeV. The other 2HDM parameters do not contribute to $a_{\mu}^{th-2HDM} $.}
\label{figmumu1}
\end{figure}

Before proceeding further, it should be mentioned that there is a very
constraining experimental bound from the muon $(g-2)_\mu$~\cite{Krawczyk:1996sm,Dedes:2001nx,Cheung:2001hz,Cheung:2003pw}. 
A detailed account on the subject can be found in Ref.~\cite{Jegerlehner:2009ry}. 
Including new physics contributions from the 2HDM will be performed
by taking off the diagrams where the SM Higgs boson takes part and adding the new 
physics contributions; that is, $\Delta a_{\mu} = 
(a_{\mu}^{exp} - a_{\mu}^{th-SM} ) +  a_{\mu}^{th-2HDM}$. 
The two most important contributions from extended models to the muon anomalous
magnetic moment are the one-loop contribution, first calculated for 
the 2HDM in Ref.~\cite{Haber:1978jt}, and 
the two-loop Barr-Zee contribution~\cite{Barr:1990vd}. 
While the one-loop contribution is proportional to $g^2_{h \mu+\mu−}$ and has therefore the SM sign, the two-loop contribution is either proportional to $g_{h\mu^+ \mu^ −} \, 
g_{h \bar{b} b}$ or to $g_{h\mu^+ \mu^ −} \, g_{h \bar{t} t}$ and can therefore have either sign in the 2HDM.
As a result, the one-loop contribution helps curing the 3$\sigma$ deviation 
relative to the SM, while
the two-loop contribution can, if sufficiently large, increase or decrease the difference
between theory and experiment.
In Fig.~\ref{figmumu1}, we present $\Delta a_{\mu} = 
(a_{\mu}^{exp} - a_{\mu}^{th-SM} ) +  a_{\mu}^{th-2HDM} $ 
as a function of $\tan \beta$ for $m_h = 120$ GeV and $m_A = 8$ GeV  
(left) and $m_A = 12$ GeV (right) in the limit $\alpha \approx \beta$.  
The other 2HDM parameters do not contribute to $a_{\mu}^{th-2HDM} $. 
It is clear from
the plots that it is possible to accommodate the contributions from the 
2HDM within the 1$\sigma$ bound even for very low $m_A$ masses.
The most striking feature is, however, the dependence on the CP-odd scalar: 
a small change in the mass can lead to a big change in its contribution to the muon anomalous magnetic moment. 

\begin{figure}[h!]
\centering
\hspace{-1.cm}
\includegraphics[width=3.2in,angle=0]{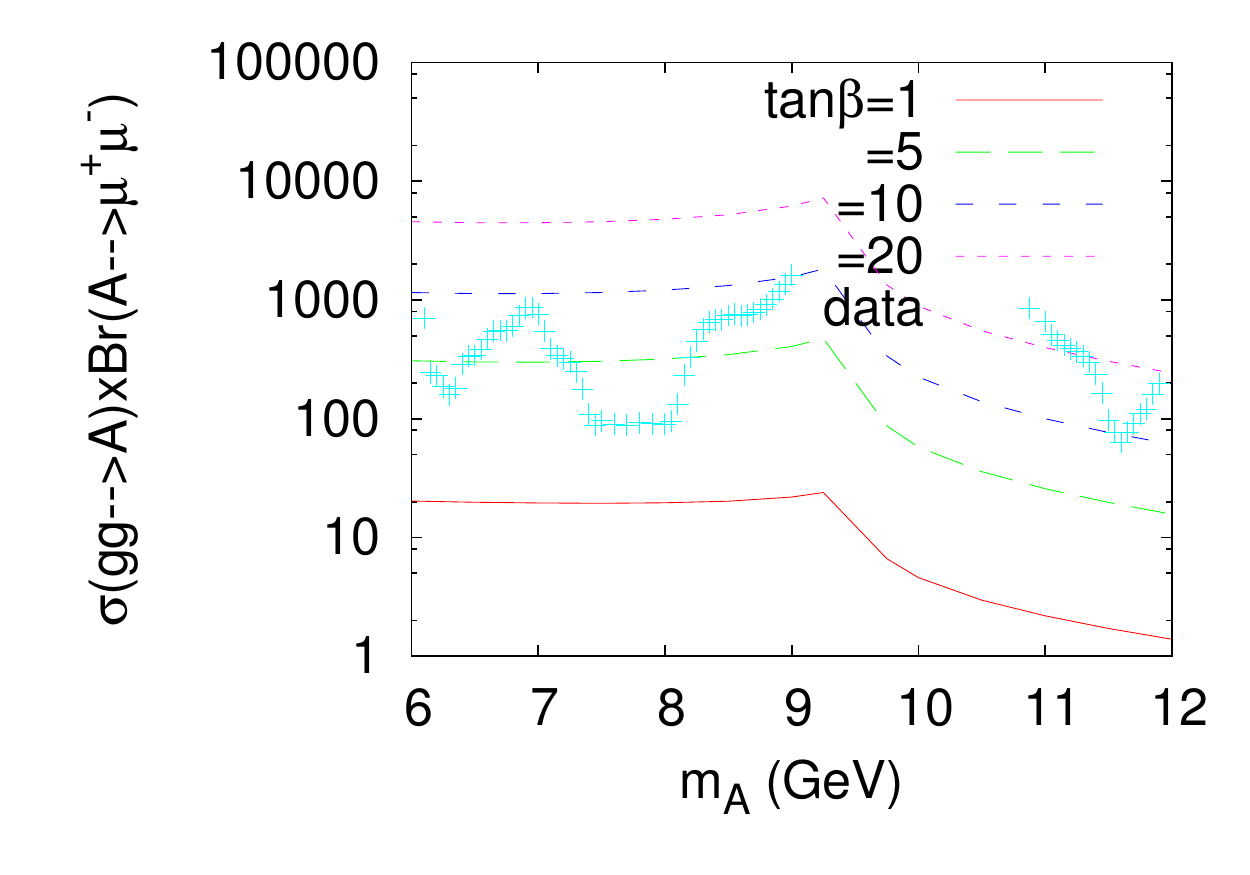}
\includegraphics[width=3.2in,angle=0]{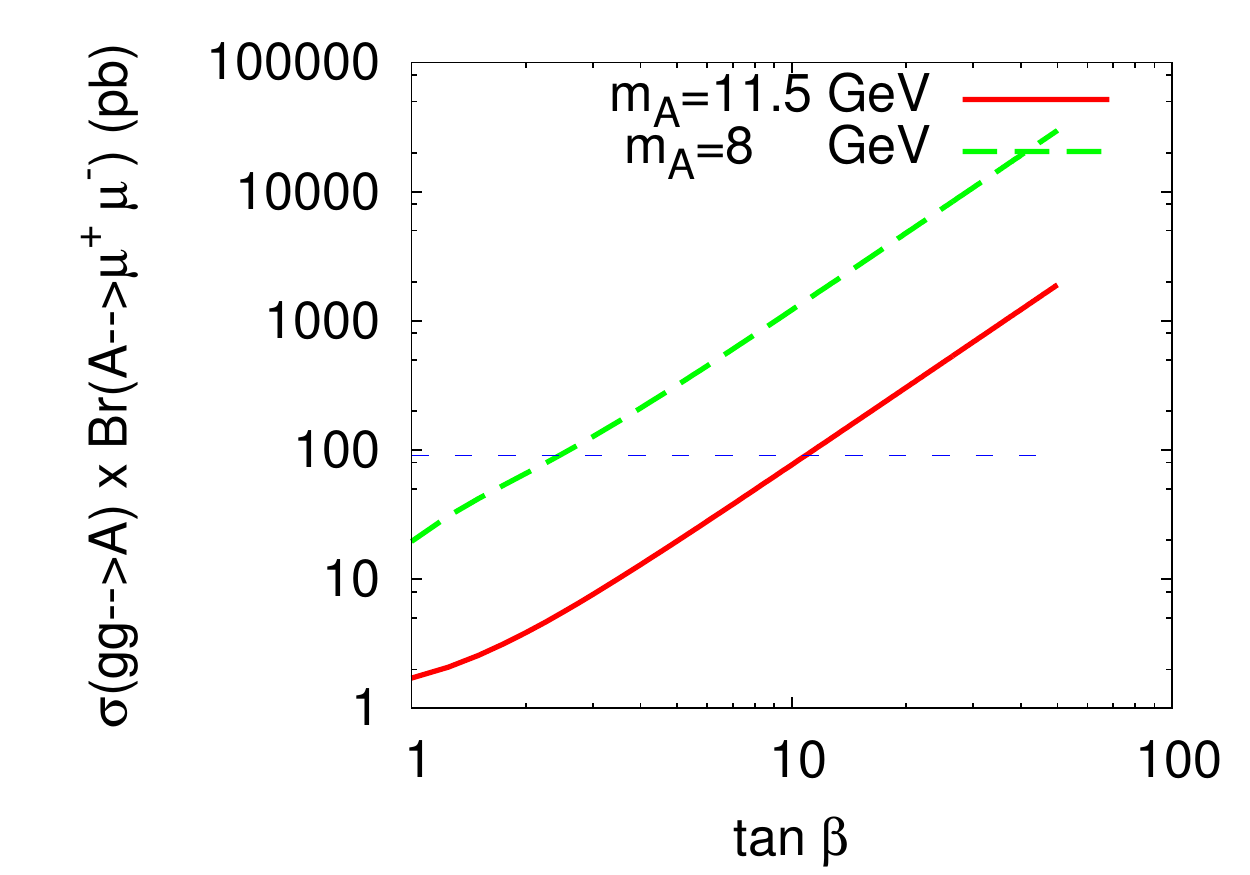}
\caption{In the left panel we show the $pp \to A \to \mu^+ \mu^-$ cross section in 2HDM-II  as a function of the pseudo-scalar mass
for several values of $\tan \beta$ together with the observed limits on $\sigma(pp \to A \to \mu^+ \mu^-)$ taken from ATLAS~\cite{ATLASmumu}. 
In the right panel we present $pp \to A \to \mu^+ \mu^-$ in 2HDM-II as a function of $\tan \beta$ for $m_A= 8$ GeV (dashed line) and $m_A= 11.5$ GeV (full line). 
The dashed horizontal line represents the experimental limit~\cite{ATLASmumu}. Limits on $\tan \beta$ are the points
where the theoretical cross section intersects with the dashed experimental line.
}
\label{figmumu2}
\end{figure}


Taking into account the Yukawa couplings, only Types II and IV are worth further exploration. However, it turns out that for model IV, all parameter space is allowed although
values of $\tan \beta  \approx 2$ are close to being excluded.
In the left panel of Fig.~\ref{figmumu2} we present the cross section for $pp \to A \to \mu^+ \mu^-$ in 2HDM-II  as a function of the pseudo-scalar mass
for several values of $\tan \beta$ together with the observed limits on $\sigma(pp \to A \to \mu^+ \mu^-)$ taken from ATLAS~\cite{ATLASmumu}. 
In the right panel of Fig.~\ref{figmumu2} we show $pp \to A \to \mu^+ \mu^-$ in 2HDM-II as a function of $\tan \beta$ for $m_A= 8$ GeV (dashed line) and $m_A= 11.5$ GeV (full line). 
The dashed horizontal line represents the experimental limit~\cite{ATLASmumu}. Limits on $\tan \beta$ are the points
where the theoretical cross section intersects with the dashed experimental line.  The lines intersect at $\tan \beta \approx 2.4$ for $m_A = 8$ GeV and $\tan \beta \approx 14.5$ for $m_A = 10.9$ GeV.  
When compared with the bounds obtained by the OPAL Collaboration, we conclude that below the $\bar{b} b$ threshold the results
are better than the ones obtained with the OPAL data.  Above the $\bar{b} b$ threshold, the results would be competitive 
with the DELPHI ones had they done the analysis for scalar masses below 12 GeV.  Since this did not happen,
results for the mass region between 11 and 12 GeV are the first ones to put a strong limit on $\tan \beta$ for Type II.
For the reasons already discussed, no limits can be extracted for the remaining Yukawa types of the model.

\section{Conclusions\label{sec:summary}}

The very recent results from the CMS and ATLAS Collaborations on 
the search for neutral Higgs bosons decaying to tau pairs at $\sqrt{s} = 7$ TeV
has prompted us to analyze its implications for CP-conserving 2HDM's
with natural flavor conservation.  From the four possible Yukawa types of the
model, we have concluded that for only two
of them the experimental constraints result in relevant exclusion 
regions in the 2HDM parameter space.  This conclusion
already takes into account the fact that other experimental constraints
have already excluded the $\tan \beta < 1$ region.

\begin{figure}[h!]
\centering
\hspace{-1.cm}
\includegraphics[width=3.2in,angle=0]{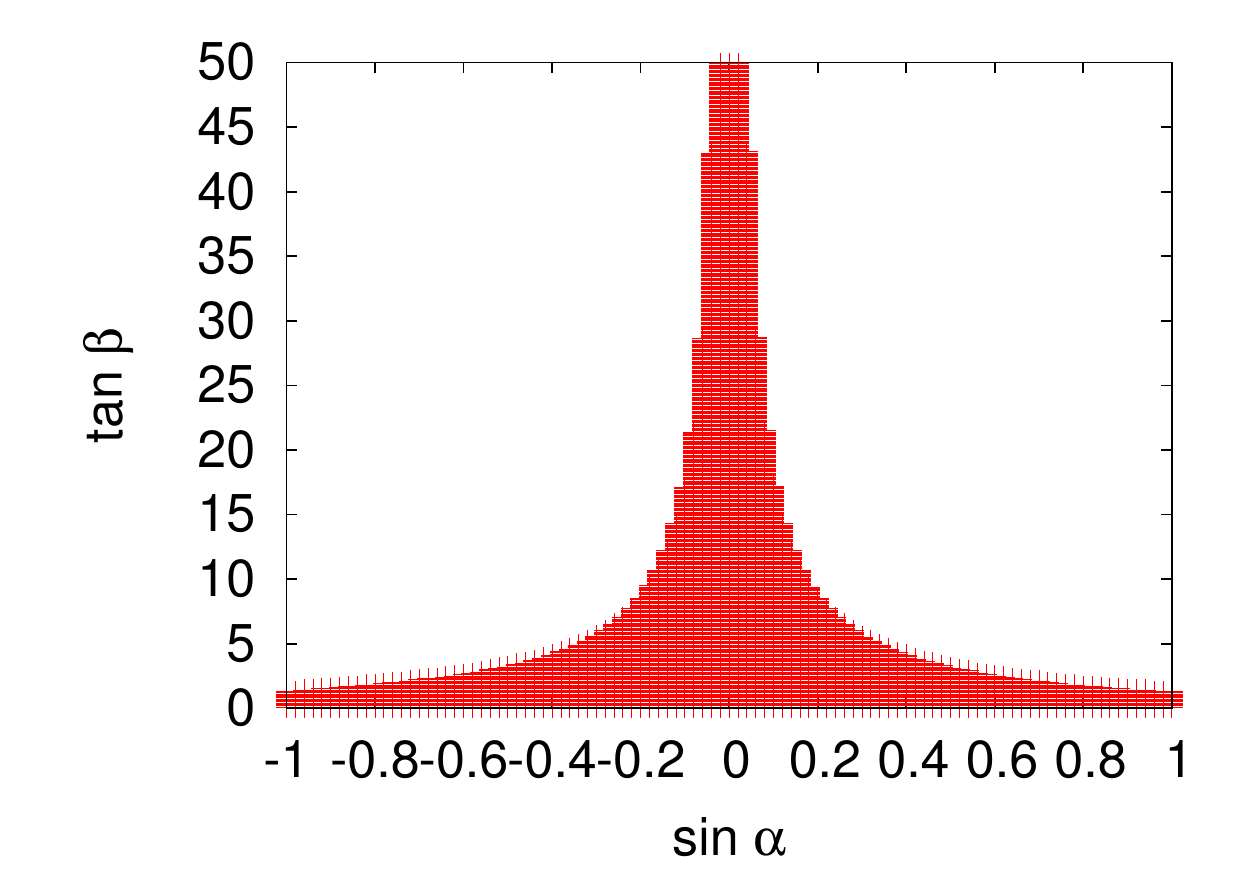}
\includegraphics[width=3.2in,angle=0]{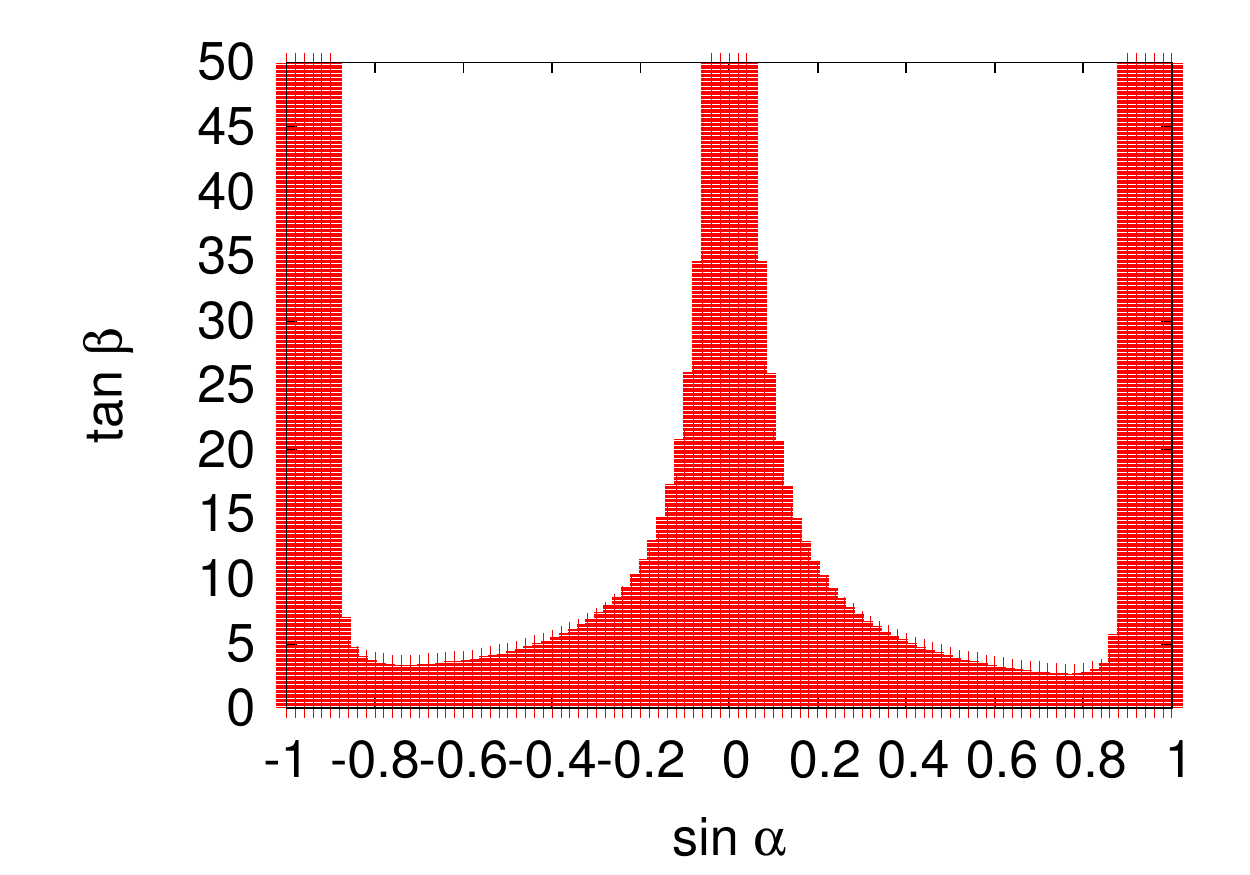}
\caption{Exclusion plots by $pp\to h\to \tau^+\tau^-$ in the
  $(\tan\beta,\sin\alpha)$ plane for $m_h=125$ GeV in 2HDM-II (left) and 
2HDM-IV (left). Red color areas are not excluded.  }
\label{figsatb125}
\end{figure}

We now summarize the results for $\tan \beta > 1$ in the four scenarios, starting with
the CP-even Higgs boson. In Type I, the branching ratios are very similar to the SM ones while the production cross section is always below the SM rate.  Model III has an increased cross
section relative to the SM, yet the branching ratio to a tau pair is much smaller due to the $\tan \beta$ dependence.  Hence, the ratio $R_{\tau \tau}$ is always below 1
for Type I and Type III.  In Fig.~\ref{figsatb125}, we show the exclusion region for $m_h=125$ GeV in the $(\tan\beta,\sin\alpha)$ plane. In the left panel for Type II, the allowed region shrinks as one moves to large values of $|\sin \alpha|$.  For $\sin \alpha \approx 0$, all values of $\tan \beta$ are allowed.  In the right panel for Type IV, both regions of small and large $|\sin \alpha|$ are
now allowed.  If we believe the 125 GeV Higgs boson hinted at by the $\gamma \gamma$ measurements, then the large values of $|\sin \alpha|$ are excluded~\cite{Ferreira:2011kt}.

\begin{figure}[h!]
\centering
\hspace{-1.cm}
\includegraphics[width=3.2in,angle=0]{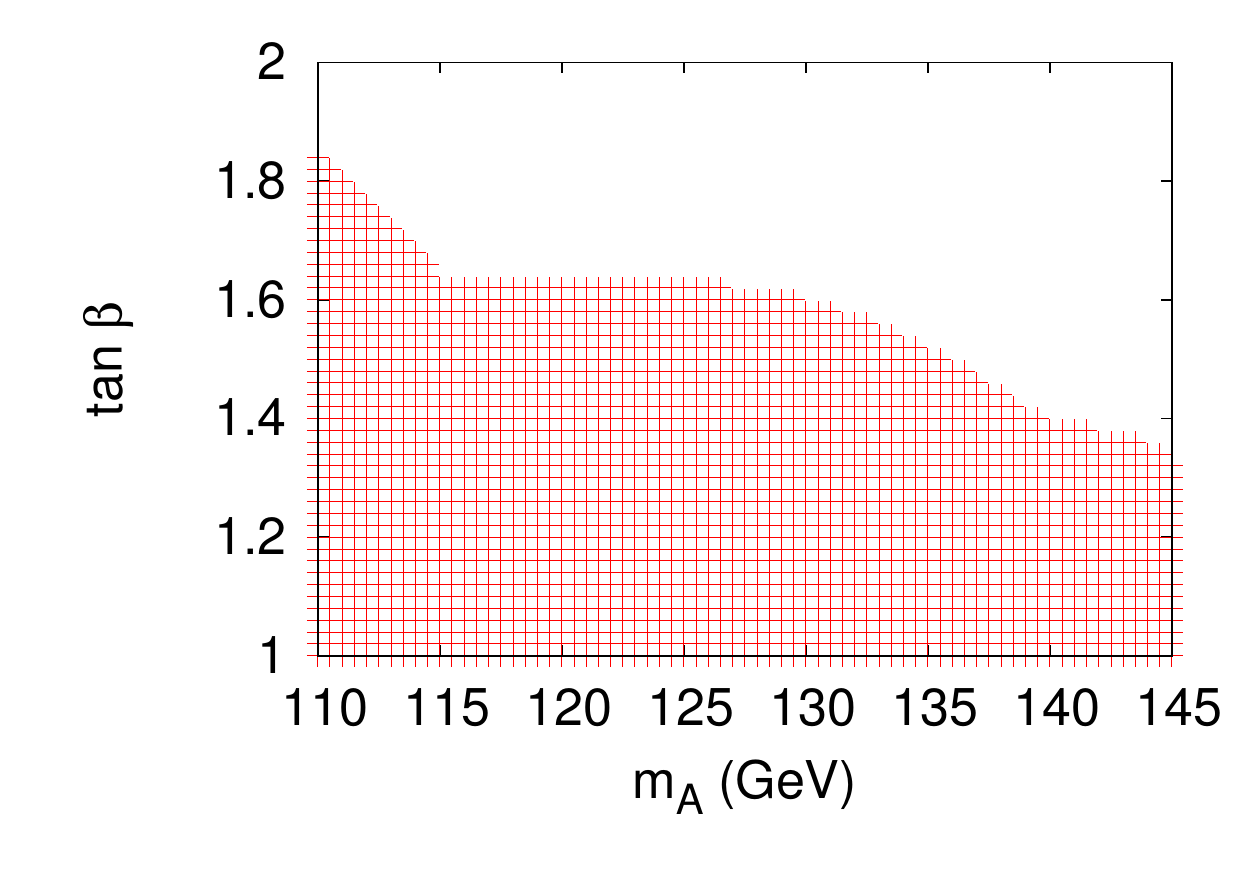}
\caption{Exclusion plot by $pp\to A\to \tau^+\tau^-$ in the $(\tan\beta,m_A)$
  plane in 2HDM-II. Values above $\tan \beta =2$ are excluded in Type II.}
\label{figtb125}
\end{figure}

The CP-odd scalar analysis leads to a similar conclusion with the bonus
that the parameter $\sin \alpha$ is not relevant.  The results
are presented in the $(\tan\beta,m_A)$ plane.  Again, for $\tan \beta > 1$
no interesting exclusion can be found for Type I, Type III and now also for Type IV.
As stated previously, values of $\tan \beta \approx 2$ are close to being
excluded in Type IV,  and more data will possible exclude some window centred
at $\tan \beta = 2$. 
In Fig.~\ref{figtb125}, we show the results for Type II where the exclusion
region is quite large.  In fact, for all
values of the CP-odd scalar mass, values of $\tan \beta >1.8 $ are definitively
excluded. Taking into account that $\tan \beta$ is already constrained
to be above 1, the present results are close to exclude all values of $\tan \beta$
if the pseudo-scalar is in the mass range considered.
The study recently performed in \cite{Burdman:2011ki} concludes that if the excess 
observed at the LHC comes from $A \to \gamma \gamma$, $\tan \beta \approx O (1)$,
which is still allowed within experimental and theoretical errors in all four models.  

\section*{Acknowledgments}
A.~A. acknowledges the support from NSC under contract \# 100-2811-M-006-008.  
C.-W.~C. is supported in part by NSC Grant No.~100-2628-M-008-003-MY4 and NCTS.  D.~K.~G. acknowledges partial support from the Department of Science and Technology, India under the grant SR/S2/HEP-12/2006.  He would also like to thank the ICTP High Energy Group, Trieste for their hospitality, where part of this work was done.  R.~S. is supported in part by the Portuguese \textit{Funda\c{c}\~{a}o para a Ci\^{e}ncia e a Tecnologia} (FCT) under contracts PTDC/FIS/117951/2010 and PEst-OE/FIS/UI0618/2011 and by an FP7 Reintegration Grant, number PERG08-GA-2010-277025.


\bibliographystyle{apsrev}
\bibliography{tau-ref-new}

\end{document}